\documentclass[12pt, a4paper]{article}
\pdfoutput=1
\usepackage[dvipdfmx]{graphicx}
\usepackage{amssymb}
\usepackage{amsmath}
\usepackage{bm}
\usepackage{multirow}
\usepackage[table]{xcolor} 
\usepackage{cite}
\usepackage{slashed}
\usepackage{subfigure}
\usepackage{epstopdf}            
\usepackage{epsfig}
\usepackage{here}
\usepackage{comment}
\usepackage{booktabs} 
\usepackage{colortbl} 
\usepackage{mathptmx} 
\usepackage{wrapfig}
\usepackage{ascmac}
\usepackage{fancybox}  
\usepackage{ulem}
\usepackage{enumerate}
\DeclareMathAlphabet{\mathcal}{OMS}{cmsy}{m}{n} 

\renewcommand{\thefootnote}{\fnsymbol{footnote}}
\numberwithin{equation}{section} 

\def\beq#1\eeq{\begin{align}#1\end{align}}
\newcommand{\ov}{\overline}

\RequirePackage{xspace}
\def\Bbar    {\kern 0.18em\overline{\kern -0.18em B}{}\xspace}
\def\Bb      {\ensuremath{\Bbar}\xspace}

\usepackage{caption}
\definecolor{BlueViolet}{rgb}{0.2, 0.00, 0.7}
\definecolor{Blue}{rgb}{0.15, 0.00, 0.9}
\definecolor{light_blue}{rgb}{0.15, 0.35, 0.95}
\definecolor{kit_green}{rgb}{0, 
0.58823 
, 0.50980 
}
\usepackage[
colorlinks=true, linkcolor=light_blue,citecolor=light_blue,urlcolor=kit_green]{hyperref}

\begin{document}
\sloppy 

\begin{titlepage}
\begin{center}

\hfill{P3H--23--036, TTP23--20}
\vskip .3in
{\Large{\bf A closer look at isodoublet vector leptoquark\\ solution to the $R_{D^{(*)}}$ anomaly }}
\vskip .3in

\makeatletter\g@addto@macro\bfseries{\boldmath}\makeatother

{\large 
Syuhei Iguro$^{\rm (a,b)}$,
Yuji Omura$^{\rm (c)}$
}
\vskip .3in

$^{\rm (a)}${\it Institute for Theoretical Particle Physics (TTP), Karlsruhe Institute of Technology (KIT),
Wolfgang-Gaede-Str.\,1, 76131 Karlsruhe, Germany}\\
\vspace{4.6pt}

$^{\rm (b)}${\it Institute for Astroparticle Physics (IAP),
Karlsruhe Institute of Technology (KIT), 
Hermann-von-Helmholtz-Platz 1, 76344 Eggenstein-Leopoldshafen, Germany}\\
\vspace{4.6pt}

$^{\rm (c)}${\it
Department of Physics, Kindai University, Higashi-Osaka, Osaka 577-8502, Japan}\\[3pt]

\> {\it E-mail:} \href{mailto:igurosyuhei@gmail.com}{igurosyuhei@gmail.com},
\href{mailto:yomura@phys.kindai.ac.jp}{yomura@phys.kindai.ac.jp},

\vskip 0.1in

\end{center}
\vskip .3in

\begin{abstract}

\noindent 
We discuss a model with a SU$(2)_L$ doublet vector leptoquark (LQ), motivated by 
the recent experimental results relating to the lepton universality of $\Bb \to D^{(*)} \tau \overline{\nu}_\tau$.
We find that scalar operators predicted by the LQ are favored to explain the deviations,
taking into account the recent LHCb result.
We investigate the extensive phenomenology of the model and conclude that $B_s\to\tau\ov\tau$, $B\to K\tau\ov\tau$, $B_u\to \tau\ov\nu_\tau$ and high-$p_T$ di-$\tau$ lepton signatures at the LHC will probe the interesting parameter region in the near future.

\end{abstract}

{{\sc Keywords:} 
Vector SU(2) doublet leptoquark,
$R_{D^{(*)}}$ anomaly,
LHC,
Flavor} 
\end{titlepage}

\setcounter{page}{1}
\renewcommand{\thefootnote}{\#\arabic{footnote}}
\setcounter{footnote}{0}

\hrule
\tableofcontents
\vskip .2in
\hrule
\vskip .4in

\section{Introduction}
\label{sec:intro}
The semi-tauonic $B$-meson decays, $\Bb \to D^{(*)} \tau \ov\nu$, have been interesting processes to measure the lepton flavor universality (LFU):
 \begin{align}
  R_D \equiv \frac{{\rm{BR}}(\Bb\rightarrow D \,\tau\, \overline\nu_\tau)}{{\rm{BR}}(\Bb\rightarrow D\, \ell\,\overline\nu_\ell)} , \qquad
  R_{D^{*}} \equiv \frac{{\rm{BR}}(\Bb\rightarrow D^{*} \tau \,\overline\nu_\tau)}{{\rm{BR}}(\Bb\rightarrow D^{*} \ell\,\overline\nu_\ell)}, 
\end{align}
where $\ell$ denotes light charged leptons.
Interestingly, deviations from the SM predictions \cite{Bordone:2019guc,Iguro:2020cpg,HFLAV:2022pwe,Bernlochner:2022ywh}\footnote{Recently the dispersive matrix approach of the form factors found the larger $R_{D^*}$ \cite{Martinelli:2021onb,Martinelli:2021myh} based on the Fermilab-MILC lattice result \cite{FermilabLattice:2021cdg} while this method produce the $3\,\sigma$ tension in the angular observable \cite{Fedele:2023ewe}.} have been reported by 
the BaBar~\cite{Lees:2012xj,Lees:2013uzd}, Belle~\cite{Huschle:2015rga,Hirose:2016wfn,Hirose:2017dxl,Abdesselam:2019dgh,Belle:2019rba} and LHCb~\cite{Aaij:2015yra,Aaij:2017uff,Aaij:2017deq,LHCb:2023zxo,LHCbRun1had} collaborations.\footnote{$R_{D^{(*)}}$ are defined by $\ell =e,~\mu$ for the BaBar/Belle and $\ell =\mu$ for the LHCb.}
Last and early this years, the LHCb collaboration reported the first result of $R_{D^*}$ along with $R_{D}$ \cite{Ciezarek:2837207} and another $R_{D^*}$ data \cite{LHCbRDst2023}, respectively.
These latest measurements are consistent with the previous world average within the uncertainty, but the resulting world average prefers larger (smaller) deviation in $R_{D}$ ($R_{D^*}$).
The current significance of the deviation is 3-4\,$\sigma$ \cite{Iguro:2022yzr} and the new physics (NP) interpretations are updated in Refs.\,\cite{Iguro:2022yzr,Aebischer:2022oqe,Fedele:2022iib,Aban:2023pgq}.\footnote{See Tab.\,6 of Ref.\,\cite{Iguro:2022yzr} for the recent summary of the situation.}
One of the significant points, compared with the previous result, is the revival of the NP interpretation with scalar operators.
The relevant interaction, in addition to the SM contribution, is
\begin{align}
 {\cal {H}}_{\rm{eff}}= 2 \sqrt2 G_FV_{cb}\biggl[ C_{S_L}O_{S_L}+C_{S_
R}O_{S_R}\biggl],
\label{eq:Ham_bctaunu}
\end{align}
with
\begin{align}
 O_{S_L} = (\overline{c}  P_Lb)(\overline{\tau} P_L \nu_{\tau}),~~~
 O_{S_R} = (\overline{c}  P_Rb)(\overline{\tau} P_L \nu_{\tau}), \label{eq:operator_bctaunu} 
\end{align}
where $P_L=(1-\gamma_5)/2$ and $P_R=(1+\gamma_5)/2$. 
The NP contribution is taken into account by the Wilson coefficients (WCs), $C_X$ ($X=S_L, \, S_R$), normalized by the SM factor of $2 \sqrt2 G_FV_{cb}$. 

It has been well known that the $B_c$ lifetime constrains the scalar interpretation \cite{Beneke:1996xe,Alonso:2016oyd,Celis:2016azn,Akeroyd:2017mhr,Blanke:2018yud,Aebischer:2021ilm}. However, the recent result makes it possible to explain the deviations at the $1\sigma$ level using the scalar operators \cite{Fedele:2022iib}.
Furthermore, the only scalar contributions enhance the polarization observable, $F_L^{D^*}$,\footnote{See Ref.~\cite{Tanaka:2012nw} for the explicit definitions.} where the SM prediction is slightly lower than the measurement \cite{Belle:2019ewo}.

A famous mediator that induces sizable semileptonic scalar contribution is a charged Higgs in a generic two Higgs doublet model (2HDM).
This possibility has been thoroughly surveyed \cite{Crivellin:2012ye,Ko:2012sv,Crivellin:2013wna,Cline:2015lqp,Crivellin:2015hha,Lee:2017kbi,Iguro:2017ysu,Iguro:2018qzf,Martinez:2018ynq,Fraser:2018aqj,Athron:2021auq,Iguro:2022uzz,Blanke:2022pjy,Fedele:2022iib,Kumar:2022rcf,Iguro:2023jju,Das:2023gfz} and it is found that sizable contribution to WC is possible only in $O_{S_L}$.
It is noted that the type-II 2HDM can contribute to $O_{S_R}$, but the contribution is not favored since the sign of $C_{S_R}$ is always negative and does not comply with data.

A leptoquark (LQ) is considered to be one of the best candidates for the $R_{D^{(*)}}$ anomaly explanation. There are three kinds of LQs widely investigated so far \cite{Angelescu:2018tyl}.
In this paper, we focus on an isodoublet vector LQ $(\text{V}_2)$ that significantly contributes to $C_{S_R}$. Recently,
the LQ is studied motivated by $R_{K^{(*)}}$ anomaly 
\cite{Kosnik:2012dj,Shaw:2018sbe,Cheung:2022zsb} and $R_{D^{(*)}}$ anomaly \cite{Sakaki:2013bfa,Cheung:2022zsb}.
The contribution of the $\rm{V}_2$ LQ to $C_{S_R}$ could be positive and solves the anomalies.
This LQ possibility is very interesting in view of the current status, but has not been well studied.\footnote{A SU(2)$_L$ singlet leptoquark U$_1$ LQ, that is predicted by the Pati-Salam model, also induces $C_{S_R}$ in general.
The $R_{D^{(*)}}$ anomaly explanation, however, would suffer from the collider bound on extra gauge boson searches, if U$_1$ is originated from the massive gauge boson in the Pati-Salam model.
$U(2)$ flavor symmetric models \cite{Barbieri:1995uv,Barbieri:1997tu,Barbieri:2011ci,Barbieri:2011fc,Barbieri:2012uh,Blankenburg:2012nx,Barbieri:2015yvd,Bordone:2018nbg,Cornella:2019hct,Fuentes-Martin:2019mun,Cornella:2021sby,FernandezNavarro:2022gst,Aebischer:2022oqe} can evade the stringent collider bounds and predict also $C_{S_R}$ accompanied by the contribution to the SM-like operator which substantially differentiates phenomenology.}
In this work, we construct an effective model with $\text{V}_2$
from the phenomenological point of view.
We study correlations between
$R_{D^{(*)}}$ and other observables, and discuss how to test this LQ possibility.

Before the recent LHCb result, the $\text{V}_2$ LQ could not explain the $R_{D^{*}}$ within $2\sigma$ \cite{Iguro:2018vqb}.
The situation, however, changes: the current minimal $\chi^2$ for $R_{D^{*}}$ becomes 3.7 with $O_{S_R}$ \cite{Iguro:2022yzr} which should be compared to $\chi^2_{\rm{SM}}=13.6$.
It would be timely to analyze the model with $\text{V}_2$.
Compared to the previous work \cite{Cheung:2022zsb}, that appeared before the LHCb results, new parts are as follows. 
First, we assign a $\tau$ number to $\rm{V}_2$. 
This assignment forbids a proton decay and suppresses many flavor violating processes. 
The underlying theory is beyond our scope, but our setup would be a guiding principle to construct a concrete model. We study relevant flavor phenomenology in this effective model.
We newly examine correlations between $R_{D^{(*)}}$ and other observables in this model, and find that $B_s\to\tau\ov\tau$ and $B\to K\tau\ov\tau$ are greatly enhanced.\footnote{Similar correlations have been scrutinized within other LQ scenarios see Refs.\,\cite{Capdevila:2017iqn,Bordone:2018nbg,Cornella:2019hct,Fuentes-Martin:2019mun,Cornella:2021sby,FernandezNavarro:2022gst,Aebischer:2022oqe} for instance. }
Second, we find that $B\to\tau\ov\nu_\tau$, that is not studied in Ref. \cite{Cheung:2022zsb}, excludes the simplest setup for the explanation of $R_{D^{(*)}}$.
We rescue the possibility by adding one more interaction.
Third, we investigate the LHC implication of the model with the help of the public tool HighPT \cite{Allwicher:2022mcg}.
We conclude that signals with two oppositely charged $\tau$ leptons in the final states will also probe the interesting parameter region in the near future.


The outline of the paper is given as follows.
In Sec.\,\ref{sec:Model} we introduce the working model for the $\rm{V}_2$ LQ and summarize the model parameters.
In Sec.\,\ref{sec:Pheno} we discuss the relevant flavor observables and investigate the phenomenology.
Then we also consider the constraint from the LHC and discuss the future prospect.
Sec.\,\ref{sec:Summary} is devoted to summary and discussion.

\section{Model setup}
\label{sec:Model}
In this section we introduce the working model and four-fermi interactions relevant to the phenomenology.

\subsection{Simplified model with the $V_2$ LQ}
\label{sec:V2model}
We shall consider an extended SM model with a SU(2)$_L$ doublet vector LQ, $\text{V}_2$.
The charge assignment of $\text{V}_2$ is (SU(3)$_c$, SU(2)$_L$, U(1)$_Y)=(\ov{3}$, 2, 5/6) and the field is described as 
\begin{eqnarray}
  \rm{V}_2 =\left(
  \begin{array}{c}
    \rm{V}_2^{4/3}\\
    \rm{V}_2^{1/3}
  \end{array}
  \right),
\label{eq:V2doublet}
\end{eqnarray}
where the electromagnetic charges of the upper and lower component are $4/3$ and $1/3$, respectively. 
This charge assignment is the same as that of a X boson in the SU(5) grand unified theory (GUT).
In this paper we do not specify the UV completion, and simply assign $\tau$ number and mass to this doublet.
As a result, a disastrous rapid proton does not occur since the di-quark coupling is forbidden by the $\tau$ number conservation.

Under this assumption the couplings between $\rm{V}_2$ and SM fermions relevant to our phenomenology are given by
\begin{align}
{\cal{L}}_{\rm{V}_2}\,=\,h_{1}^{ij}(\ov{d^{C}_i}\gamma_\mu  P_L L^{b}_j)\epsilon^{ab}{\rm{V}}_2^{\mu,a}
+h_{2}^{ij}(\ov{Q^{C,a}_i}\gamma_\mu  P_R e_j)\epsilon^{ab}{\rm{V}}_2^{\mu,b}
+h_3^{ij}(\ov{Q^{C}_i}\gamma_\mu P_R u_j){\rm{V}}_2^{\mu *}+\rm{h.c.}\nonumber\\
\end{align}
where indices $i,\,j$ and $a,\,b$ are labels of flavor and SU(2)$_L$.
We work on the down-quark basis.
This choice is phenomenologically conservative since flavor changing neutral currents involving light down-type are suppressed at tree level.
It is noted that within a $\mathcal{O}(1)$\,TeV LQ scenarios 1-loop induced processes, e.g. meson mixing, is important although an UV completion is required to evaluate the correction. One possible underlying theory will be discussed in Sec. \ref{sec:Summary}.
It is noted that those interactions change the fermion number by 2 units: $|\Delta F|=2$ where $F=3B+L$ and, $B$ and $L$ are baryon and lepton numbers, respectively.
By assigning the $\tau$ number to $\rm{V}_2$ we can eliminate $h_3$ that triggers a dangerous proton decay \cite{Frampton:1989fu,Frampton:1990hz,Frampton:1991ay}.\footnote{$\tau$-flavored $\rm{U}_1$ LQ is discussed in Ref.\,\cite{Bernigaud:2021fwn}.}
Thanks to the $\tau$ charge assignment, the structure of the interaction is described as 
 \begin{eqnarray}
  h_1^{ij} =\left(
  \begin{array}{ccc}
    0 &0 &h_1^{13}\\
    0 &0&h_1^{23}\\
    0 &0 &h_1^{33}\\
  \end{array}
  \right),~~~h_2^{ij} =\left(
  \begin{array}{ccc}
    0 &0&h_2^{13}\\
    0 &0&h_2^{23}\\
    0 &0&h_2^{33}\\
  \end{array}
  \right),~~~ h_3=0.
\label{eq:coupling_structure}
\end{eqnarray}
Assuming that those elements are real, we consider flavor and collider phenomenology in the next section. Now, the terms in ${\cal{L}}_{\rm{V}_2}$ are decomposed as  
\begin{align}
{{\cal{L}}_{\rm{V}_2}}\,=\,&+h_{1}^{i3}(\ov{d^{C}_i}\gamma_\mu P_L \tau){\rm{V}}_2^{4/3,\mu}
-h_{1}^{i3}(\ov{d^{C}_i}\gamma_\mu  P_L \nu_{\tau}){{{\rm{V}}_2^{1/3,\mu}}}\nonumber\\
&-h_{2}^{i3}(\ov{d^{C}_i}\gamma_\mu  P_R \tau){\rm{V}_2^{4/3,\mu}}
+h_{2}^{i3}(\ov{u^{C}_i}\gamma_\mu  P_R \tau){\rm{V}_2^{1/3,\mu}}+\rm{h.c.}.
\label{Eq:fermoin_VQ}
\end{align}
The mass eigenstates are given by replacing as 
$(u_L,\,d_L) \to (V_Q^\dagger u_L,\,d_L)$
\begin{align}
{{\cal{L}}_{\rm{V}_2}}\,=\,&+h_{1}^{i3}(\ov{d^{C}_i}\gamma_\mu P_L \tau){\rm{V}_2^{4/3,\mu}}-h_{1}^{i3}(\ov{d^{C}_i}\gamma_\mu  P_L \nu_{\tau})\rm{V}_2^{1/3,\mu}\nonumber\\
&-h_{2}^{i3}(\ov{d^{C}_i}\gamma_\mu  P_R \tau)\rm{V}_2^{4/3,\mu}+(V_Q^*h_{2})^{i3}(\ov{u^{C}_i}\gamma_\mu  P_R \tau)\rm{V}_2^{1/3,\mu}+\rm{h.c.},
\label{Eq:fermoin_VQ2}
\end{align}
where $V_Q$ denote 
Cabbibo-Kobayashi-Maskawa matrix \cite{Cabibbo:1963yz,Kobayashi:1973fv}.

\subsection{Four-fermi couplings}
\label{sec:fourfermi}

The interactions in Eq.\,(\ref{Eq:fermoin_VQ2}) contribute to the semileptonic operators through $V^{4/3}$ and $V^{1/3}$ exchanges:
\begin{align}
{\mathcal L}_{NDSL}=&-\frac{h_1^{i3}h_1^{k3*} }{m_{\rm{V}_2}^2} \left(\ov{d}_k \gamma_\mu P_R d_i\right) \left(\ov{\tau} \gamma^\mu P_L \tau \right) 
    -\frac{h_1^{i3}h_1^{k3*} }{m_{\rm{V}_2}^2} \left(\ov{d}_k \gamma_\mu P_R d_i\right) \left(\ov{\nu}_\tau \gamma^\mu P_L \nu_\tau \right) \nonumber\\
    &-\frac{h_2^{i3}h_2^{k3*} }{m_{\rm{V}_2}^2} \left(\ov{d}_k \gamma_\mu P_L d_i\right) \left(\ov{\tau} \gamma^\mu P_R \tau \right) 
    -\frac{(V_Q^* h_2)^{i3}(V_Q^* h_2)^{k3*} }{m_{\rm{V}_2}^2} \left(\ov{u}_k \gamma_\mu P_L u_i\right) \left(\ov{\tau} \gamma^\mu P_R \tau \right) \nonumber\\
    &+\frac{2h_1^{i3}h_2^{k3*} }{m_{\rm{V}_2}^2} \left(\ov{d}_k  P_R d_i\right) \left(\ov{\tau} P_L \tau \right) 
    +\frac{2h_1^{i3}(V_Q^* h_2)^{k3*} }{m_{\rm{V}_2}^2} \left(\ov{u}_k  P_R d_i\right) \left(\ov{\tau} P_L \nu_\tau \right)  \nonumber\\
    &+\frac{2h_2^{i3}h_1^{k3*} }{m_{\rm{V}_2}^2} \left(\ov{d}_k  P_L d_i\right) \left(\ov{\tau} P_R \tau \right) 
    +\frac{2(V_Q^* h_2)^{i3}h_1^{k3*} }{m_{\rm{V}_2}^2} \left(\ov{d}_k  P_L u_i\right) \left(\ov{\nu}_\tau P_R \tau \right),
    \label{EQ:NDSL}
\end{align}
where the masses of $\rm{V}_2^{4/3}$ and $\rm{V}_2^{1/3}$ are assumed to be degenerate and $m_{\rm{V}_2}$ denotes the LQ mass.
We categorize these four-fermi interactions, based on the induced processes:
\begin{enumerate}[(i)]
\item down type neutral current ($\tau$),
\item down type neutral current ($\nu_\tau$),
\item up type neutral current,
\item charged current.
\end{enumerate}
Our main goal of this paper is to find the correlation between $R_{D^{(*)}}$, to which $h_1^{33}\times h_2^{23*}$ dominantly contributes, and other observables.
We introduce the following hierarchical coupling structure,
\begin{eqnarray}
  h_1^{ij} =\left(
  \begin{array}{ccc}
   ~0~&~0~&~\epsilon\\
   ~0~&~0~&~\epsilon\\
   ~0~&~0~&~\mathcal{O}(1)\\
  \end{array}
  \right),~~~h_2^{ij} =\left(
  \begin{array}{ccc}
   ~0~&~0~&~\epsilon\\
   ~0~&~0~&~\mathcal{O}(1)\\
   ~0~&~0~&~\epsilon\\
  \end{array}
  \right),
\label{eq:coupling_structure_2}
\end{eqnarray}
where $\epsilon$ is a small dimensionless parameter.

It is noted that $|h_{1,2}^{i3}|^2$ does not trigger lepton flavor violating processes, although it is important in collider phenomenology as we will see later.
At $\mathcal{O}(\epsilon^0)$, we focus on the combination of $h_1^{33}h_2^{23*}$, that contributes to categories (i) and (iv).
At $\mathcal{O}(\epsilon^1)$, we have 8 combinations that involve all of four categories.
Those 9 combinations and the relevant flavor processes are summarised in Tab.\,\ref{Tab:ModelNRC}.
Below, we summarize the parameterizations in the four categories.

\begin{table}[p]
\begin{center}
\scalebox{1.1}{
  \begin{tabular}{c|c|c} \hline
   Coupling product & $\rm{V}^{4/3}_2$ & $\rm{V}^{1/3}_2$ \\ \hline \hline 
  \multirow{3}{*}{$h_1^{33}\times h_2^{23*}$} & (i)~~$b\to s\tau\bar{\tau}$ & (iv)~~$b\to c\tau\ov\nu_{\tau}$  \\ 
   &\multirow{2}{*}{$B_s\to \tau\ov\tau$,~$B\to K\tau\ov\tau$}&$\Bb \to D^{(*)}\tau\ov\nu_\tau$\\ 
    & & $B_c\to\tau\nu_\tau$,~$B_u\to\tau\ov\nu_\tau$\\ \hline\hline
  \multirow{3}{*}{$h_1^{33}\times h_2^{13*}$} & (i)~~$b\to d\tau\bar{\tau}$ & (iv)~~$b\to u\tau\ov\nu_{\tau}$  \\ 
   &\multirow{2}{*}{$B_d\to \tau\ov\tau$,~$B\to\pi\tau\ov\tau$}&$\Bb \to D^{(*)}\tau\ov\nu_\tau$\\
    & & $B_u\to\tau\nu_\tau$,~$B\to\pi\tau\ov\nu_\tau$\\ \hline
   \multirow{2}{*}{$h_1^{33}\times h_2^{33*}$} &(i)~~$b\bar{b}\to\tau\ov{\tau}$ & (iv)~~$t\to b\tau\ov\nu_\tau$ \\ 
   &\sout{$\Upsilon(nS)\to\tau\ov\tau$}&---\\ \hline
   \multirow{2}{*}{$h_1^{33}\times h_1^{13*}$} &(i)~~$b\to d\tau\ov\tau$ & (ii)~~$b\to d\nu_\tau\ov\nu_\tau$  \\
   &$B_d\to\tau\ov\tau$, $B\to \pi\tau\ov\tau$&\sout{$B_d\to\nu\ov\nu_\tau$},~$B\to\pi\nu_\tau\ov\nu_\tau$\\ \hline
   \multirow{2}{*}{{$h_1^{33}$}$\times h_1^{23*}$} &(i)~~$b\to s\tau\ov\tau$ & (ii)~~$b\to s\nu_\tau\ov\nu_\tau$  \\ 
   &$B_s\to\tau\ov\tau$,~$B\to K\tau\ov\tau$&\sout{$B_s\to \nu_\tau\ov\nu_\tau$},~$B\to K \nu_\tau\ov\nu_\tau$\\ \hline
   \multirow{2}{*}{$h_1^{13}\times h_2^{23*}$} &(i)~~$s\to d\tau\ov\tau$ & (iv)~~$c\to d\tau\ov\nu_\tau$  \\ 
   &---&{$D_d\to\tau\ov\nu_\tau$}\\ \hline
   \multirow{2}{*}{$h_1^{23}\times h_2^{23*}$} &(i)~~$s\ov{s}\to\tau\ov\tau$ & (iv)~~$c\to s \tau\ov\nu_\tau$  \\ 
   &---& {$D_s\to\tau\ov\nu_\tau$} \\ \hline
   \multirow{2}{*}{$h_2^{13}\times h_2^{23*}$} &(i)~~$s\to d\tau\ov\tau$ & (iii)~~$c\to u\tau\ov\tau$  \\ 
   &---&---\\ \hline
   \multirow{2}{*}{$h_2^{33}\times$$h_2^{23*}$} &(i)~~$b\to s \tau\ov\tau$&(iii)~~$t\to c \tau\ov\tau$  \\
   &$B_s\to \tau\ov \tau$,~$B\to K \tau\ov\tau$ & ---  \\ \hline
   \end{tabular}
  }
  \caption{Summary table for the relevant flavor processes.
  In the first row we list up the category and parton level processes and if it exists mesonic in the second row.
  Processes with the strikethrough are prohibited by the symmetry argument or suppressed by the neutrino mass.
   }
  \label{Tab:ModelNRC}
\end{center}   
\vspace{-.45cm}
\end{table}

\subsection*{(i) Down type neutral current ($\tau$)}
In the categories (i), the induced operators are 
\begin{align}
\mathcal{H}_{\rm eff}^{\tau} =-\frac{\alpha G_F V_{td_i}V_{td_k}^*}{\sqrt{2}\pi}& \Big ( C_S^{ki}O_S^{ki} + C_S^{ki \prime}O_S^{ki \prime }+C_P^{ki}O_P^{ki} + C_P^{ki \prime}O_P^{ki\prime}  \nonumber \\
& +C_9^{ki}O_9^{ki} + C_9^{ki \prime}O_9^{ki \prime }+C_{10}^{ki}O_{10}^{ki} + C_{10}^{ki \prime}O_{10}^{ki\prime} \Big ) +{\rm{h.c.,}}
\end{align}
where  
\begin{align}
O_S^{ki}&=(\ov{d}_k P_R d_i)(\ov{\tau} \tau),~~~~O_{P}^{ki}=(\ov{d}_k P_R d_i)(\ov{\tau} \gamma_5\tau),\nonumber\\
O_9^{ki}&=(\ov{d}_k\gamma_\mu P_L d_i)(\ov{\tau}\gamma^\mu \tau), ~~~~O_{10}^{ki}=(\ov{d}_k \gamma_\mu P_L d_i)(\ov{\tau}\gamma^\mu \gamma_5\tau),
\label{eq:ND_SL_ope}
\end{align}
and the primed operators are obtained by exchanging $P_L\leftrightarrow P_R$.
Matching onto the WCs at the LQ scale is
\begin{align}
C_S^{ki}&=-C_P^{ki}=\frac{\sqrt{2} \pi }{\alpha G_F V_{td_i} V_{td_k}^* }\frac{h_1^{i3}h_2^{k3*}}{m_{\rm{V}_2}^2},~~~
C_S^{ki\prime}=C_P^{ki\prime}=\frac{\sqrt{2} \pi }{\alpha G_F V_{td_i} V_{td_k}^* }\frac{h_2^{i3}h_1^{k3*}}{m_{\rm{V}_2}^2},\nonumber\\
C_9^{ki}&=C_{10}^{ki}=-\frac{ \pi }{\sqrt{2}\alpha G_F V_{td_i} V_{td_k}^* }\frac{h_2^{i3}h_2^{k3*}}{m_{\rm{V}_2}^2},~~~
C_9^{ki\prime}=-C_{10}^{ki\prime}=-\frac{ \pi }{\sqrt{2}\alpha G_F V_{td_i} V_{td_k}^* }\frac{h_1^{i3}h_1^{k3*}}{m_{\rm{V}_2}^2}.
\label{eq:ND_SL_ope2}
\end{align}
The relative factor of 2 and sign difference between scalar and vector operators come from the Fierz identities. 
It is noted that $h_1\times h_1$ and $h_2\times h_2$ contribute to vector operators while $h_1\times h_2$ contributes to scalar operators. As we will see below, we find that the scalar operators are correlated with the charged current, while vector operators are independent of $R_{D^{(*)}}$ because of the structure.
We note that there is a chirality enhancement in purely leptonic meson decays with the contribution of the scalar operators.

\subsection*{(ii) Down type neutral current ($\nu_\tau$)}
The induced operators involving $\nu_\tau$ are
\begin{align}
\mathcal{H}_{\rm eff}^{\nu} = - \frac{ \sqrt{2} G_F \alpha}{\pi} V_{t d_i} V^*_{td_k} C_R^{ki} \left( \ov{d}_i \gamma^\mu P_R d_k\right) \left( \ov\nu_\tau \gamma_\mu P_L \nu_\tau \right),
\end{align}
where 
\begin{align}
    C_{R}^{ki}=-\frac{h_1^{i3} h_1^{k3*}}{m_{\rm{V}_2}^2}\frac{\pi}{ \sqrt{2} G_F \alpha V_{td_i} V^*_{td_k}}.
\end{align}
The combination of $h_1^{i3}h_1^{k3*}$ contributes to this category mediated by the $V^{1/3}$ LQ and only vector operators are generated. 
As a result $M_1\to \nu_\tau\ov\nu_\tau$ process is suppressed by the neutrino mass and negligible in our setup where $M$ denotes a meson.
Therefore we focus on $M_1\to M_2 \nu_\tau \ov\nu_\tau$.

\subsection*{ (iii) Up type neutral current}
The $h_2 \times h_2$ combination only gives the operators involving $\tau$ and up-type quarks.
$h_2^{23}\times h_2^{33*}$ and $h_2^{23}\times h_2^{13*}$ induce $t \ov c \tau\ov\tau$ and $c \ov u \tau\ov\tau$ vector operators respectively.
Regarding the latter interaction, it is difficult to obtain the constraint at the tree level in flavor physics because of the heavy $\tau$ mass with respect to the charm mass.
Although $t \to c \tau\ov\tau$ transition is kinematically allowed, the experimental sensitivity to BR($t \to c \tau\ov\tau$) is several orders away from the prediction even at the high luminosity (HL) LHC \cite{Kim:2018oih}.
Therefore we will not discuss the physics induced by those terms below.

\subsection*{(iv) Charged current}
Finally, we discuss the charged current involving $\tau$.
This interaction contributes to $R_{D^{(*)}}$, and described by the $h_1\times h_2^*$ combination.
The resulting semitauonic scalar operator is 
\begin{align}
 {\mathcal H}_{\rm{CSR}}= 
 2 \sqrt 2 G_F V_{ki} C_{S_R}^{u_kd_i}(\overline{u_k} P_R d_i)(\overline{\tau} P_L \nu_{\tau}) .
 \label{Eq:CSL}
\end{align} 
where the coefficient at the LQ scale, $\mu_{\rm LQ}$, is evaluated as
\begin{align}
C_{S_R}^{u_kd_i}(\mu_{\rm{LQ}})=-\sum_{n}\frac{V_Q^{kn}}{\sqrt{2}G_F V_Q^{ki}}\frac{h_1^{i3}h_2^{n3*}}{m_{\rm{V}_2}^2}.
\label{eq:CSR_general}
\end{align}
We note that $i=3,\,k=2$ corresponds to Eq.\,(\ref{eq:Ham_bctaunu}).
The operator triggers $M_1^-\to \tau\ov\nu_\tau$ and $M_1\to M_2 \tau\ov\nu_\tau$ decays.
It is noted that the former again receives the chirality enhancement while the enhancement in the case with $M_1=D_{(s)}$ is moderate due to $m_{D_{(s)}} \simeq m_\tau$.

\section{Phenomenology}
\label{sec:Pheno}
In this section, we discuss the phenomenology in this model, assuming the LQ couplings are aligned as in Eq. (\ref{eq:coupling_structure_2}).
In Sec.\,\ref{sec:minimal_flavor}, we study the processes where 
our predictions are not suppressed by $\epsilon$ nor CKM $\lambda$ in the Wolfenstein parameterization\cite{Wolfenstein:1983yz}.
The LHC phenomenology of $\mathcal{O}(\epsilon^0)$ will be given in Sec.\,\ref{sec:LHC}.
In Sec.\,\ref{sec:next_to_minimal_flavor}, we discuss our predictions at $\mathcal{O}(\epsilon^0)$ and $\mathcal{O}(\lambda)$.
Finally the $\mathcal{O}(\epsilon)$ phenomenology is given in Sec.\,\ref{sec:Oepsilon}.

\subsection{Flavor phenomenology at $\mathcal{O}(\epsilon^0)$ and $\mathcal{O}(\lambda^0)$ }
\label{sec:minimal_flavor}
First of all, we consider the $b\to c\tau\bar{\nu}_\tau$ transition corresponding to the category (iv).
As discussed above, the semileptonic charged current is generated by the ${\rm{V}}^{1/3}_2$ exchange, and it is proportional to 
$h_1^{33} \times h_2^{23*}$.
The induced operator, $O_{S_R}$, at the LQ scale is evaluated as
\begin{align}
C_{S_R}^{cb}(\mu_{\rm{LQ}})\,\simeq\,0.18\left(\frac{2\,{\rm{TeV}}}{m_{\rm{V}_2}}\right)^2\left(\frac{h_1^{33}h_2^{23*}}{-0.5}\right).
\label{eq:eq:CSRcb}
\end{align}
We adopt the generic formula given in Ref.\,\cite{Iguro:2022yzr} for the prediction of $R_{D^{(*)}}$.
It is known that the imaginary part of $C_{S_R}$ is not helpful to fit the current $R_{D^{(*)}}$ result and hence we assume those couplings to be real.
The constraint on WC from high-$p_T$ di-$\tau$ search, which we see later in this section, is almost independent of the LQ scale. 
As a benchmark, we set the LQ mass to $m_{\rm{V}_2}=2\,$TeV.
To connect the coefficient to the $B$ meson scale, $\mu_b=4.2$\,GeV, we use the renormalization group evolution (RGE) for the dimension-six operators at the QCD next-to-leading and the electroweak leading orders including the top-quark threshold corrections \cite{Jenkins:2013wua,Alonso:2013hga,Gonzalez-Alonso:2017iyc,Aebischer:2017gaw}.
We also include the QCD one-loop matching corrections \cite{Aebischer:2018acj}.
As a result we approximately obtain 
\begin{align}
C_{S_R}(\mu_b)=2.0\,C_{S_R}(\mu_{\rm LQ}).
\end{align}
Thanks to the LHCb downward (upward) shift of $R_{D^*}$ ($R_D$), we find that  $C_{S_R}$ can explain the anomaly within $2\sigma$.

Next we consider the $B_s \to \tau\ov\tau$ decay, that is predicted by
the operators in the category (i). This process is correlated with $R_{D^{(*)}}$ in our model.
Using the operators in Eq.\,(\ref{eq:ND_SL_ope}) the branching ratio is given as
\begin{align}
    {\rm{BR}}(B_s\to\tau\ov\tau)=&\frac{\tau_{B_{s}}f_{B_{s}}^2 m_{B_{s}} G_F^2 m_\tau^2 \alpha^2 |V_{tb} V_{ts}|^2}{16 \pi^3} \sqrt{1-\frac{4m_\tau^2}{m^2_{B_{s}}}}\times \nonumber\\
    &\left \{ \left |(C_{10}^{sb}-C_{10}^{sb\prime})+\frac{m^2_{B_{s}}}{2m_b m_\tau}(C_P^{sb}-C_P^{sb\prime}) \right|^2+\frac{m_{B_{s}}^4}{4m_b^2 m_\tau^2} \left( 1- \frac{4m_\tau^2}{m_{B_{s}}^2}\right) \left |C_S^{sb}-C_S^{sb\prime} \right|^2 \right \}.
\label{eq:Bstotautau}
\end{align}
We note that the coefficient in the SM is estimated as
$C_{10}^{sb,{\rm{SM}}}(\mu_b)=-4.3$ \cite{Bobeth:1999mk,Huber:2005ig}.
In our model, the scalar semileptonic operator is induced at the tree level, so that the leptonic meson decay has the chirality enhancement.
Currently the LHCb with Run\,1 data sets the leading upper limit on the decay as \cite{LHCb:2017myy} 
\begin{align}
{\rm{BR}}(B_s\to\tau\bar{\tau})\le6.8\times10^{-3}.
\end{align}
The future prospects of the Run\,3 and the HL LHC are estimated in Ref. \cite{LHCb:2018roe}: compared to the current bound the sensitivities will be improved by factor of 5 and 13, respectively.

The coupling product also contributes to $B\to K \tau\ov\tau$.
The current limit announced by the BaBar collaboration is BR$(B\to K \tau\ov\tau)\le 2.25\times 10^{-3}$ \cite{BaBar:2016wgb} where the SM prediction is $1.4\times 10^{-7}$ \cite{Capdevila:2017iqn}.
The relevant formula is given as \cite{Aebischer:2022oqe}
\begin{align}
    \frac{\rm{BR} (B\to K\tau\ov\tau)}{ \rm{BR} (B\to K\tau\ov\tau)_{\rm{SM}}}\simeq\left(1+0.17{\rm{Re}}[C_P^{sb}]+0.14{\rm{Re}}[C_S^{sb}] +0.06|C_S^{sb}|^2+0.06 |C_P^{sb}|^2\right).
    \label{eq:BtoKtautau}
\end{align}
The Belle II experiment with 5\,ab$^{-1}$ of data \cite{Kou:2018nap} will be sensitive to BR$(B\to K\tau\ov\tau)=6.5\times 10^{-5}$.
\begin{figure}[t]
\begin{center}
 \includegraphics[width=0.5 \textwidth]{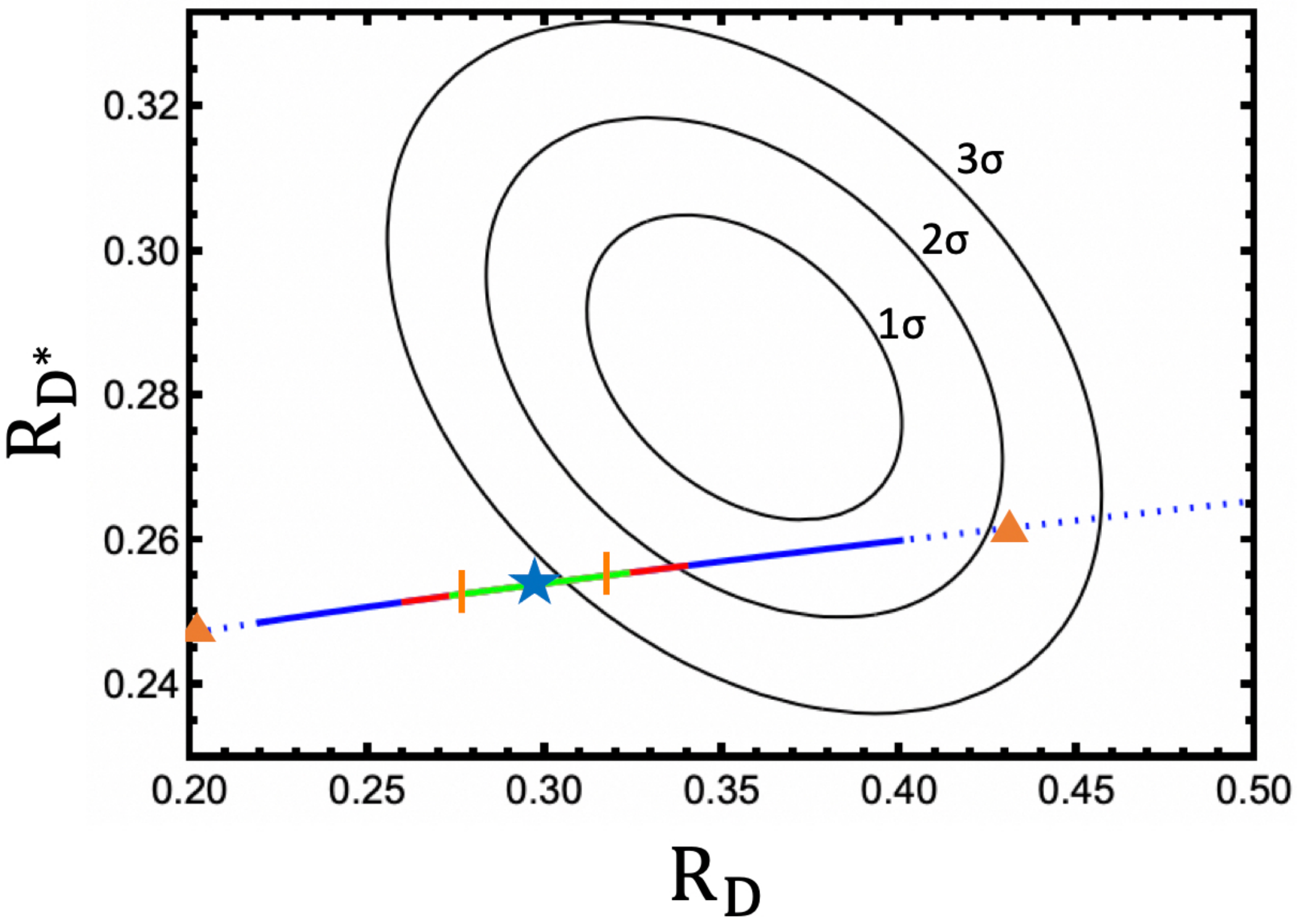}
\end{center}
 \caption{
Correlation among $R_{D^{(*)}}$, BR($B_s\to \tau\ov \tau$) and BR($B\to K\tau\ov \tau$) is shown.
Colored lines are prediction of $\rm{V}_2$ LQ model. 
The star mark corresponds to the SM prediction.
Blue dotted lines are excluded by the $B_s\to\tau\ov\tau$ measurement.
Blue solid lines (red solid lines) will be probed with $B_s\to\tau\ov\tau$ at Run\,3 (HL LHC).
The current constraint and near future prospect of the $B\to K\tau\ov\tau$ measurement are shown in orange.
Region between triangles are currently allowed and the gap between the triangle and vertical bar will be probed with Belle II early data of 5\,ab$^{-1}$. 
$1,\,2,\,3\,\sigma$ contours for $R_{D^{(*)}}$ are shown in black.
  \label{fig:RD_Bstautau}
 }
\end{figure}

It is noted that other LQ, $\text{S}_1$, $\text{R}_2$ and U(2) flavored $\text{U}_1$ do not contribute to a single scalar operator.
Given that $R_{D^{(*)}}$ anomaly is explained by $O_{S_R}$ in the $\rm{V}_2$ LQ scenario, the coupling product, $h_1^{33}\times h_2^{23}$, should be sizable. 
As shown in Eqs.\,(\ref{eq:ND_SL_ope2}), (\ref{eq:Bstotautau}) and (\ref{eq:BtoKtautau}), the sizable $h_1^{33}\times h_2^{23}$ enhances BR($B_s\to\tau\ov\tau$) and BR($B\to K\tau\ov\tau$) so that they are key predictions to test this model.
This correlation has not been pointed out in the previous works to our best knowledge.
In Fig.\,\ref{fig:RD_Bstautau}, we show the correlation among $R_{D^{(*)}}$,  BR($B_s\to\tau\ov\tau$) and BR($B\to K\tau\ov\tau$).
The colored horizontal line is the model prediction of $R_{D^{(*)}}$.
Blue dashed region is excluded by the current $B_s\to\tau\ov\tau$ constraint.
Blue and red solid lines are expected to be probed at the Run\,3 and the HL LHC.
The green would not be uncovered.
We see that the Run\,3 and the HL LHC data will be the interesting probe of the 2\,$\sigma$ region.
For the SM prediction that is depicted by a star symbol, we adopt the latest HFLAV2023 average of $R_{D}=0.298$ and $R_{D^{*}}=0.254$.\footnote{It is noted if we rely on the Lattice predictions of $R_{D}=0.299$ and $R_{D^{*}}=0.265$ \cite{MILC:2015uhg,FermilabLattice:2021cdg,Fedele:2022iib}, where $R_{D^*}$ is shifted by 0.01, the model prediction goes through the 1\,$\sigma$ region. 
On the other hand, if we adopt $R_{D}=0.290$ and $R_{D^{*}}=0.248$ \cite{Iguro:2020cpg} where the form factor is also fitted also with full angular data from the Belle \cite{Belle:2017rcc,Belle:2018ezy}, the $\rm{V}_2$ prediction contour goes though the 2\,$\sigma$ region.}
We also see the current constraint of $B\to K\tau\ov\tau$ in Fig. \ref{fig:RD_Bstautau}.
The region between the two orange triangles satisfies the $B\to K\tau\ov\tau$ constraint. 
It is found that the constraint is weaker than that from $B_s\to \tau\ov\tau$.
The Belle II with 5\,ab$^{-1}$ of data will probe the region between triangle and vertical orange line.  
Therefore $B\to K\tau\ov\tau$ will probe the interesting parameter region in near future.

The same semitauonic scalar operator which again corresponds to the category (iv) largely enhances $B_c\to\tau\ov\nu_\tau$ branching ratio.
Although $R_{D^*}$ is not largely deviated from the SM prediction, $R_{D^*}$ and BR$(B_c\to\tau\ov\nu_\tau)$ has a correlation:
\begin{align}
 \label{eq:Bc}
{\rm{BR}}(B_c\to \tau\overline\nu_\tau) ={\rm{BR}}(B_c\to \tau\overline\nu_\tau)_{\rm SM}\left| 1 + \frac{m_{B_c}^2}{m_b+m_c}C_{S_R}^{cb} \right|^2 ,    
\end{align}
where BR$(B_c\to \tau\overline\nu_\tau)_{\rm SM} \simeq 0.022$ \cite{Iguro:2022yzr}.
In the numerical evaluation, $m_b=4.18$\,GeV and $m_c(m_b)=0.92$\,GeV are used.
The upper bound of the coupling product from  BR$(B_s\to\tau\ov\tau)$ indirectly set that of ${\rm{BR}} (B_c\to\tau\ov\nu_\tau)$ as
\begin{align}
{\rm{BR}} (B_c\to\tau\ov\nu) \le 8\,\%.
\end{align}
This satisfies the current conservative limit, ${\rm{BR}} (B_c\to\tau\ov\nu_\tau) \lesssim 60\,\%$ \cite{Blanke:2018yud} while future lepton colliders can test the SM prediction at $\mathcal{O}(1)\,\%$ accuracy \cite{Zheng:2020ult,Amhis:2021cfy,Fedele:2023gyi}.

\subsection{LHC phenomenology at $\mathcal{O}(\epsilon^0)$ }
\label{sec:LHC}
\begin{figure}[t]
\begin{center}
 \includegraphics[width=0.3 \textwidth]{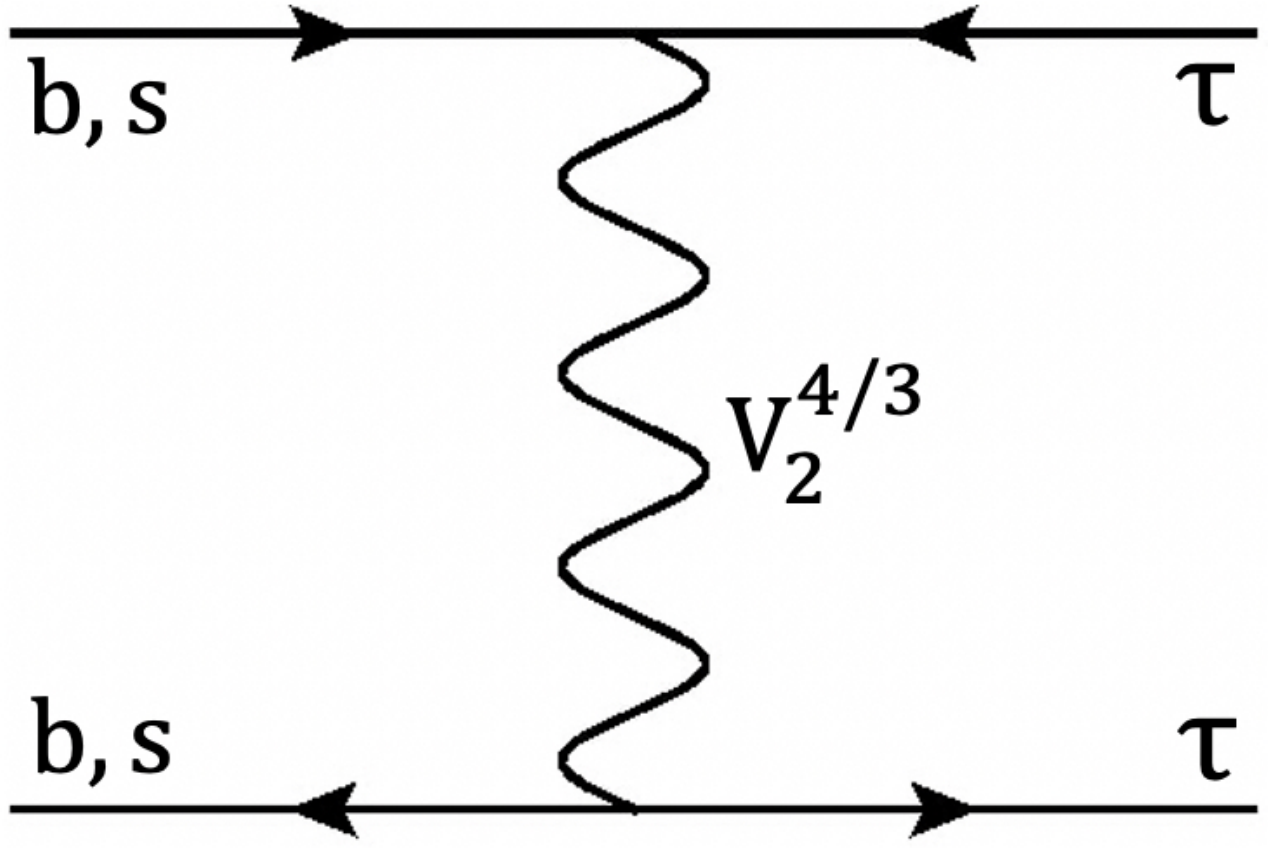}~~~~~~~~~~
  \includegraphics[width=0.3 \textwidth]{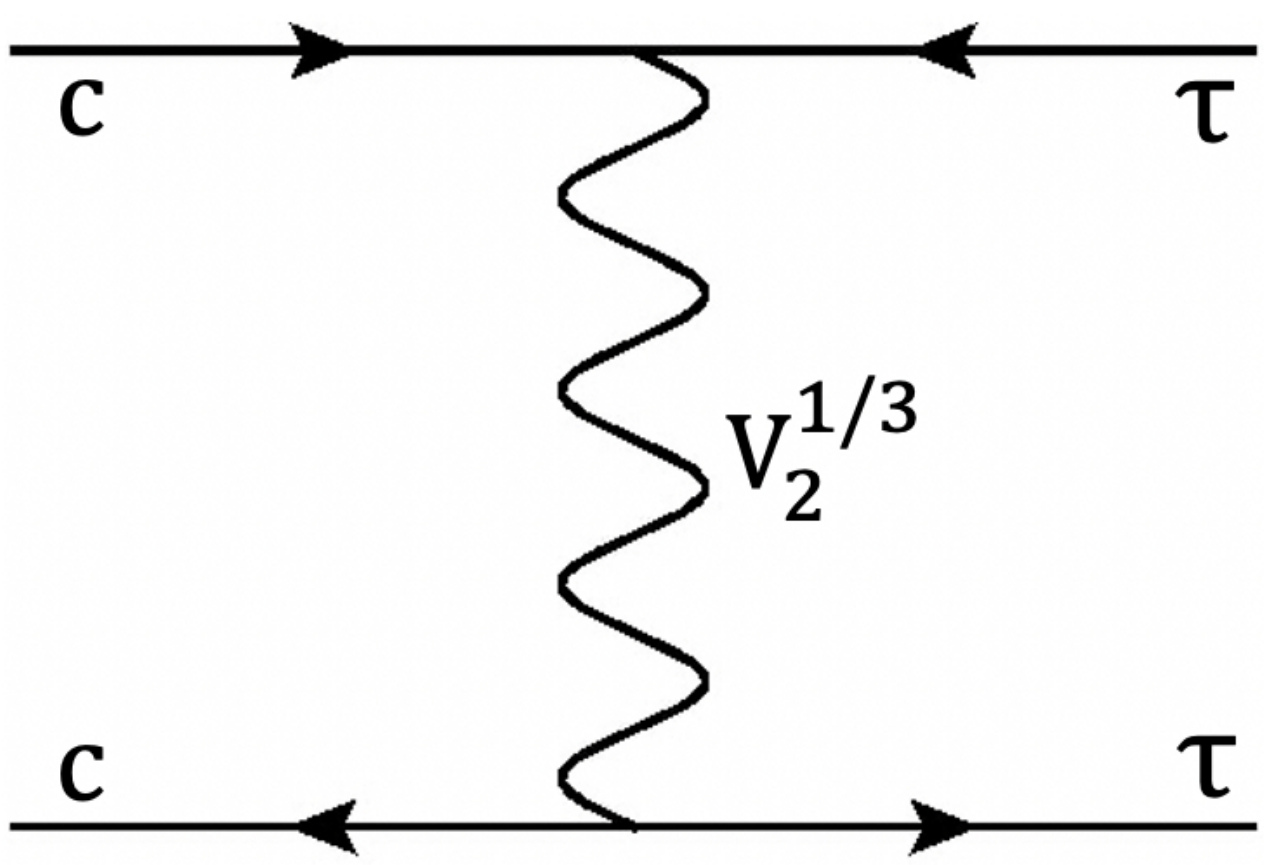}
\end{center}
 \caption{
The contributing Feynman diagrams for $\tau\ov\tau$ final state at the LHC.
Both $\rm{V}^{4/3}_2$ (left) and $\rm{V}^{1/3}_2$ (right) contribute to the high-$p_T$ signature.
  \label{fig:diagrams}
 }
\end{figure}

Since the LQ has a TeV-scale mass, the direct search at the LHC is a powerful tool to probe the scenario. $\rm{V}_2$ is pair-produced by the strong interaction at the hadron collider. Depending on the subsequent decays, we can set the lower limit on the $\rm{V}_2$ mass.
The sizable $h_{1}^{33}$ and $h_{1}^{23}$ respectively lead the following decays:  $\rm{V}^{1/3}_2\to b\nu_\tau$, $\rm{V}^{4/3}_2\to b\tau$, $\rm{V}^{1/3}_2\to c\tau$ and $\rm{V}^{4/3}_2\to s\tau$.
The LQ mass has been directly constrained as $M_\text{LQ} \gtrsim 1.5\,\text{TeV}$ from the searches for the LQ pair-production~\cite{Sirunyan:2018vhk,Aaboud:2019bye,Aad:2021rrh,ATLAS:2021jyv}. 
Furthermore, it is known that the high-$p_T$ region is important to prove the new physics scenario that explains the $R_{D^{(*)}}$ anomaly \cite{Faroughy:2016osc,Iguro:2018fni,Mandal:2018kau,Greljo:2018tzh,Altmannshofer:2017poe,Iguro:2017ysu,Abdullah:2018ets,Marzocca:2020ueu,Bhaskar:2021pml,Iguro:2020keo,Endo:2021lhi}.
In our model, $\rm{V}^{4/3}_2$ also contributes to the di-$\tau$ final state, so that the searches for di-$\tau$ with high-$p_T$ \cite{CMS:2022goy,ATLAS:2020zms} provide the better probe than the $\tau\nu$ searches studied in Refs. \cite{CMS:2015hmx,Aaboud:2018vgh,Sirunyan:2018lbg,ATLAS:2021bjk}. 
See Fig.\,\ref{fig:diagrams} for the contributing Feynman diagrams.
We study the bounds from the  di-$\tau$ and mono-$\tau$ signatures at the LHC.
We constructed the $\chi^2$ function based on the high-$p_T$ bins of Refs.\,\cite{ATLAS:2020zms,ATLAS:2021bjk}
using HighPT \cite{Allwicher:2022mcg} as a function of couplings and draw the upper bounds on the LQ couplings where $m_{\rm{V}_2}=2\,$TeV is fixed.
It is found that the mediator mass dependence in di-$\tau$ final state is mild in terms of the four-fermi interactions.
We note that the study of the interplay between $R_{D^{(*)}}$ and collider physics in this model has not been done before.
\begin{figure}[t]
\begin{center}
 \includegraphics[width=0.45 \textwidth]{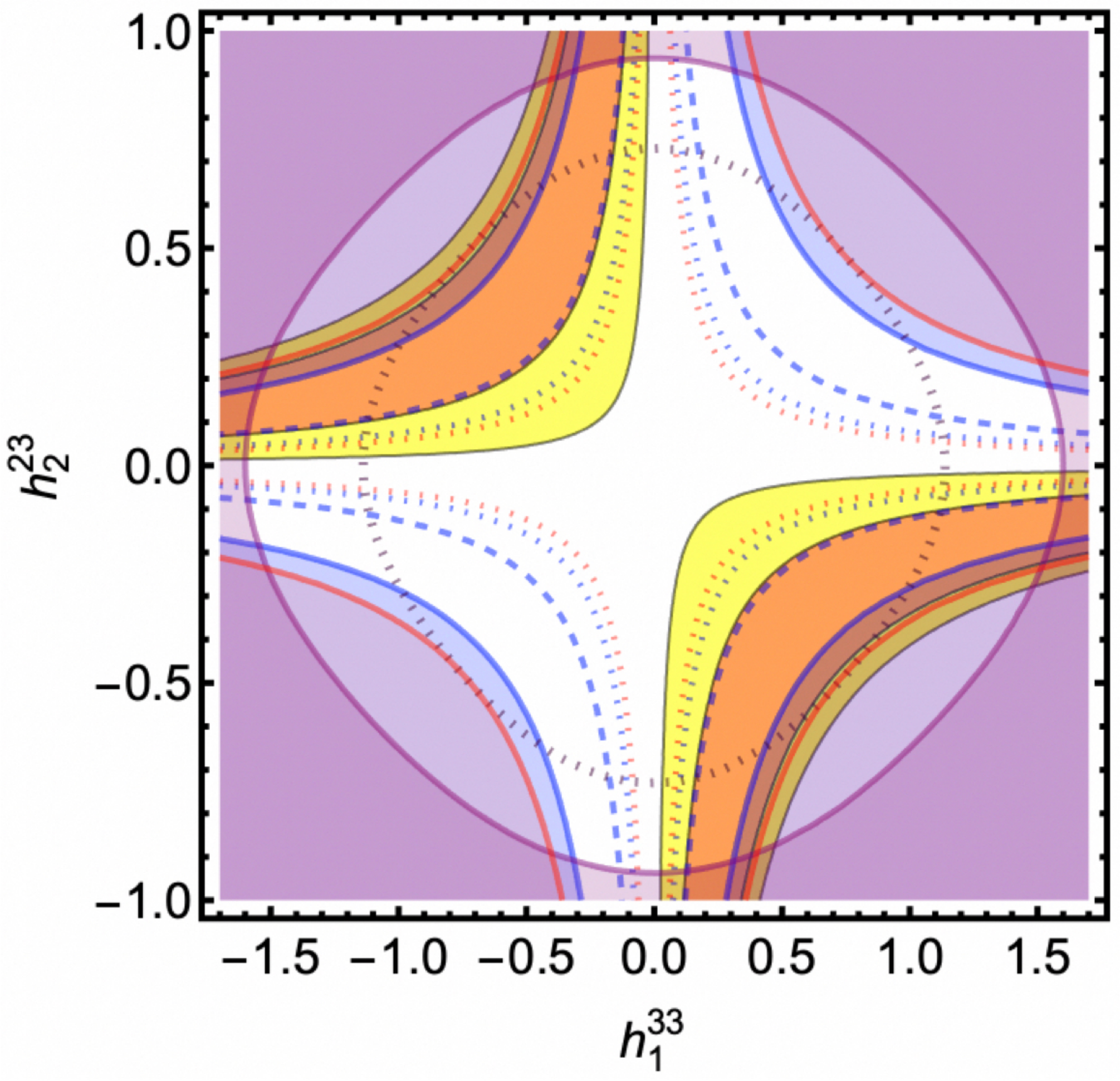}
\end{center}
 \caption{
$R_{D^{(*)}}$ favored region, constraint from $B_s\to\tau\ov\tau$ and di-$\tau$ searches at the LHC are shown on $h_1^{33}$ vs. $h_2^{23}$.
We fixed the $\rm{V}_2$ LQ mass to 2\,TeV.
Orange and yellow regions correspond to $\chi^2\le6.18$ (orange) and $\chi^2\le11.83$ (yellow) for the $R_{D^{(*)}}$ data, respectively.
Blue shaded region is excluded by the current $B_s\to\tau\ov\tau$ and dashed and dotted contours denote the future prospect for the Run 3 and HL LHC.
Similarly red shaded region is excluded by the current $B\to K\tau\ov\tau$ and dotted contours denote the future prospect for the early Belle II of 5\,ab$^{-1}$.
Purple shaded region is also excluded by the high-$p_T$ di-$\tau$ searches at the LHC.
The future projection is shown in the dotted contour.
  \label{fig:h133_h223}
 }
\end{figure}

In Fig.\,\ref{fig:h133_h223} solid and dotted purple lines show the current bound and future prospect of the LHC experiment, respectively.
The shaded region is excluded.
We overlaid the current constraint and future sensitivity of the Run\,3 and the HL LHC from $B_s\to\tau\ov\tau$ with solid, dashed and dotted blue lines.
We also show the current constraint and early Belle II sensitivity of $B\to K \tau\ov\tau$ with solid and dotted red lines.
The regions favored by $R_{D^{(*)}}$ are shown in orange and yellow: $\chi^2\le6.18$ (orange) and $\chi^2\le11.83$ (yellow).
We see that the Run\,3 $B_s\to\tau\ov\tau$, early Belle II $B\to K\tau\ov\tau$ and high-$p_T$ tail at the HL LHC will test the whole orange region and hence probe the remaining interesting parameter region.

We briefly summarize the difference in the prediction of the other LQ scenarios:
\begin{itemize}
\item {BR($B_s\to\tau\ov\tau$) and BR($B\to K\tau\ov\tau$) are largely enhanced, while, for instance, it is not in the $\rm{S}_1$ LQ case.
Although $\rm{R}_2$ and U(2) flavored $\rm{U}_1$ LQ enhance BR($B_s\to\tau\ov\tau$) and BR($B\to K\tau\ov\tau$), the former (latter) has $C_T$ ($C_{V_L}$) contribution in $R_{D^{(*)}}$ too. 
Therefore the degree of the enhancement is milder for the other LQs.
Since the coupling strength to explain the deviation is larger than the $\rm{U}_1$ LQ model, we can test this scenario with smaller amount of the data.} 
\item {Furthermore, as is shown in Ref.\,\cite{Iguro:2018vqb}, polarization observables in $\Bb\to D^{(*)} \tau\ov\nu_\tau$ are helpful to distinguish those scenarios.
Especially $\tau$ polarization will be a key observable.}
\item {The larger signal rate in di-$\tau$ is predicted at the LHC, compared to the $\rm{U}_1$ LQ.
This is because that the larger couplings are necessary to explain $R_{D^{(*)}}$, and both $\rm{V}^{4/3}_2$ and $\rm{V}^{1/3}_2$ contribute to the processes.
Therefore the LHC data in high-$p_T$ di-$\tau$ channel will be very important to probe the model.}
\end{itemize}

\subsection{Flavor phenomenology at $\mathcal{O}(\epsilon^0)$ and $\mathcal{O}(\lambda^1)$ }
\label{sec:next_to_minimal_flavor}
\begin{figure}[t]
\begin{center}
 \includegraphics[width=0.5 \textwidth]{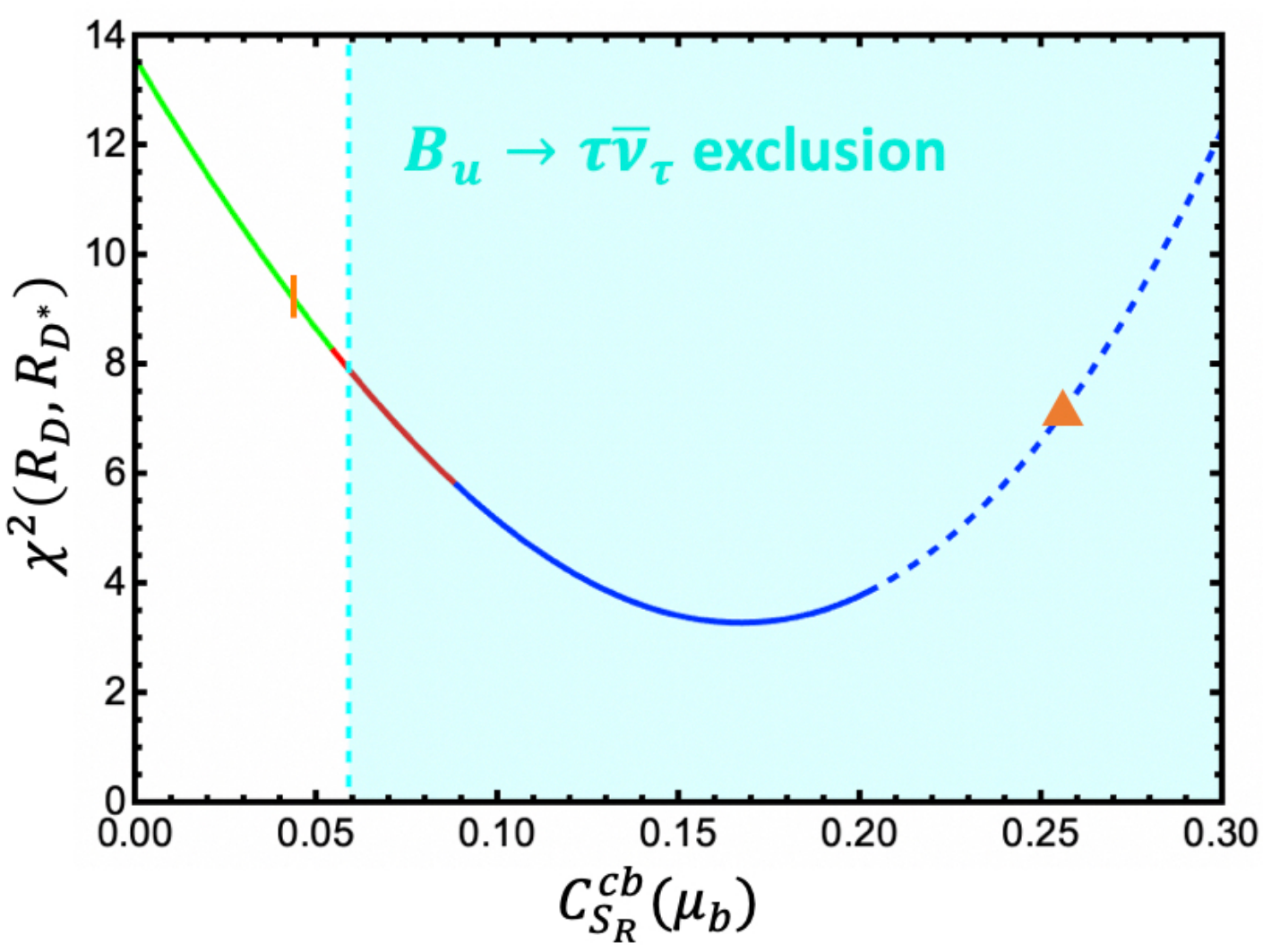}
\end{center}
 \caption{
$\chi^2(R_D,\,R_{D^{*}})$ and $R_{B_u}$ as a function of $C_{S_R}^{cb}(\mu_b)$.
$C_{S_R}^{cb}=(0.88+2.45i)C_{S_R}^{ub}$ is fixed for $R_{B_u}$. The meaning of the color is the same as in Fig.\,\ref{fig:RD_Bstautau}.
The blue dashed line is excluded by the current bound from $B_s\to\tau\ov\tau$.
The shaded cyan region is excluded by the current $B_u\to\tau\ov \nu_\tau$.
\label{fig:RD_Btaunu}
}
\end{figure}

The leptonic $B$ meson decay, $B_u\to \tau\ov\nu_\tau$, also constrain our model.\footnote{ This constraint has not been pointed out in the previous work \cite{Cheung:2022zsb}.}
This decay is enhanced by the scalar operator, although it is suppressed by an off-diagonal CKM element.
Similar to the $B_c$ decay, we can derive the numerical formula as 
\begin{align}
 \label{eq:Bu}
{\rm{BR}}(B_u\to \tau\overline\nu_\tau) ={\rm{BR}}(B_u\to \tau\overline\nu_\tau)_{\rm SM}\left| 1 + 3.75 C^{ub}_{S_R} \right|^2.
\end{align}
The SM prediction is estimated as BR$(B_u\to \tau\ov\nu_\tau)_{\rm SM}\simeq 0.95 \times 10^{-4}$ with $|V_{ub}|=0.409\times 10^{-2}$.\footnote{This is the average of inclusive and exclusive $V_{ub}$.}
$C_{S_R}^{ub}$ at the LQ scale is estimated as
\begin{align}
    C_{S_R}^{ub}(\mu_{\rm{LQ}})\simeq(0.15+0.44i)\left(\frac{2\,{\rm{TeV}}}{m_{\rm{V}_2}}\right)^2\left(\frac{h_1^{33}h_2^{23*}}{-0.5}\right).
    \label{eq:eq:CSRub}
\end{align}
It is noted that the coefficient in Eq.\,(\ref{eq:eq:CSRub}) is bigger than that of Eq.\,(\ref{eq:eq:CSRcb}) because of the factor of $(V_{cb} V_{us})/V_{ub}=0.84+2.46i$.

The current experimental world average is BR$(B_u\to \tau\overline\nu_\tau)=(1.09\pm0.24)\times 10^{-4}$ \cite{PDG2022}.
There is a notorious discrepancy between the inclusive and the exclusive determinations:
$|V_{ub}|_{\rm{inc}}=4.25\,(1\pm0.07)\times 10^{-3}$ and  $|V_{ub}|_{\rm{exc}}=3.70\,(1\pm0.04)\times 10^{-3}$. 
Therefore, we assign $14\,\%$ uncertainty to the SM amplitude.
On the other hand, the experimental result has $22\%$ uncertainty.
Combining those uncertainties at $2\sigma$, we allow $70\%$ uncertainty and set the following criteria:
\begin{align}
R_{B_u}=\frac{{\rm{BR}}(B_u\to \tau\ov\nu_\tau)}{{\rm{BR}}(B_u\to \tau\ov\nu_\tau)_{\rm{SM}}}\le1.7.
\label{eq:Butaunu_current}
\end{align}
It is noted that the following observables, 
\begin{align}
    R_{pl}=\frac{{\rm{BR}}(B_u\to\tau\ov\nu_\tau)}{{\rm{BR}}(B_u\to\mu\ov\nu_\mu)},~~~R_{ps}=\frac{\Gamma(B_u\to\tau\ov\nu_\tau)}{\Gamma(B_d\to\pi^+l\ov\nu_l)}
\end{align}
are free from $V_{ub}$ and useful to test the NP \cite{Tanaka:2016ijq}. The corresponding SM predictions are $R_{pl}^{\rm{SM}}=222$ and $R_{ps}^{\rm{SM}}=0.54\pm0.04$.\footnote{The large part of the uncertainty comes from $B\to \pi$ transition form factor \cite{Tanaka:2016ijq}.}
The current experimental constraint is given as $R_{ps}^{\rm{exp}}=0.73\pm0.14$ while $R_{ps}$ is not measured due to the large uncertainty in BR$(B_u\to\mu\ov\nu_\mu)$ \cite{Kou:2018nap}.
At $2\,\sigma$ level, this leads to $R_{B_u}\lesssim 2$, that is weaker than the constraint in Eq.\,(\ref{eq:Butaunu_current}).

The Belle II with 50\,ab$^{-1}$ of the data will measure  $R_{pl}$ and $R_{ps}$ at $12\,\%$ and $7\,\%$ at $1\,\sigma$.
Even if we adopt the current theoretical uncertainty for $R_{ps}$ to be conservative, we obtain the similar uncertainty.
It is noted that in our model, the modification of the denominator mode is negligible and hence the uncertainty of the ratio corresponds to the sensitivity to BR$(B_u\to\tau\ov\nu_\tau)$. 

In Fig.\,\ref{fig:RD_Btaunu} we show $\chi^2$ for $R_{D}$ and $R_{D^*}$.
The meaning of color and style of lines are the same as in Fig.\,\ref{fig:RD_Bstautau}.
The current conservative exclusion of Eq.\,(\ref{eq:Butaunu_current}) is shown in cyan region.
It is seen that currently $B_u\to\tau\ov\nu_\tau$ is more sensitive to $h_1^{33}\times h_2^{23*}$ than $B_s\to\tau\ov\tau$ and $B\to K\tau\ov\tau$. 
This figure clearly shows that the interesting parameter space is already excluded by the current result of $R_{B_u}$. 
This bound, however, can be relaxed by sizable other LQ couplings, as discussed in Sec.\,\ref{sec:Oepsilon}.

\subsection{$\mathcal{O} (\epsilon)$ phenomenology}
\label{sec:Oepsilon}
\begin{figure}[t]
\begin{center}
 \includegraphics[width=0.3 \textwidth]{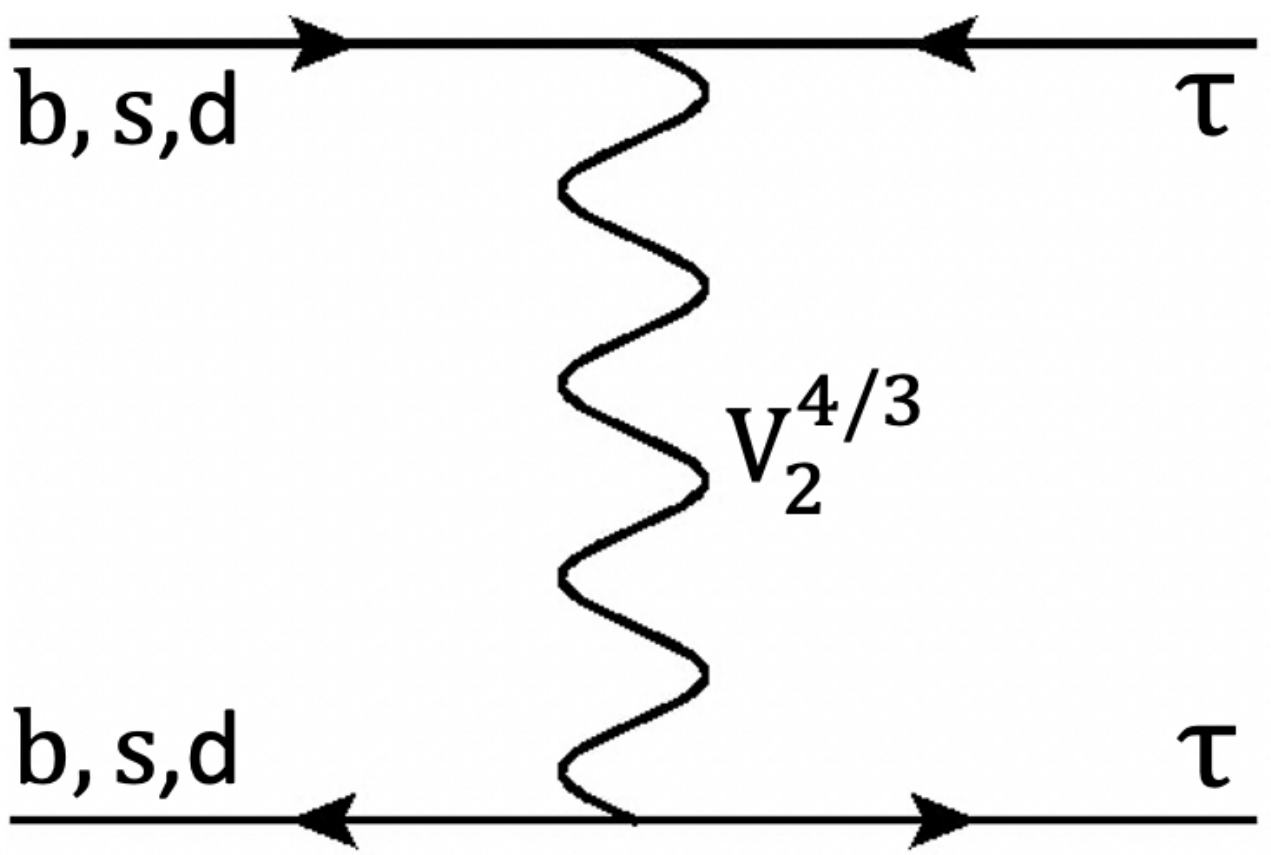}~~~~~~~~~~
  \includegraphics[width=0.3 \textwidth]{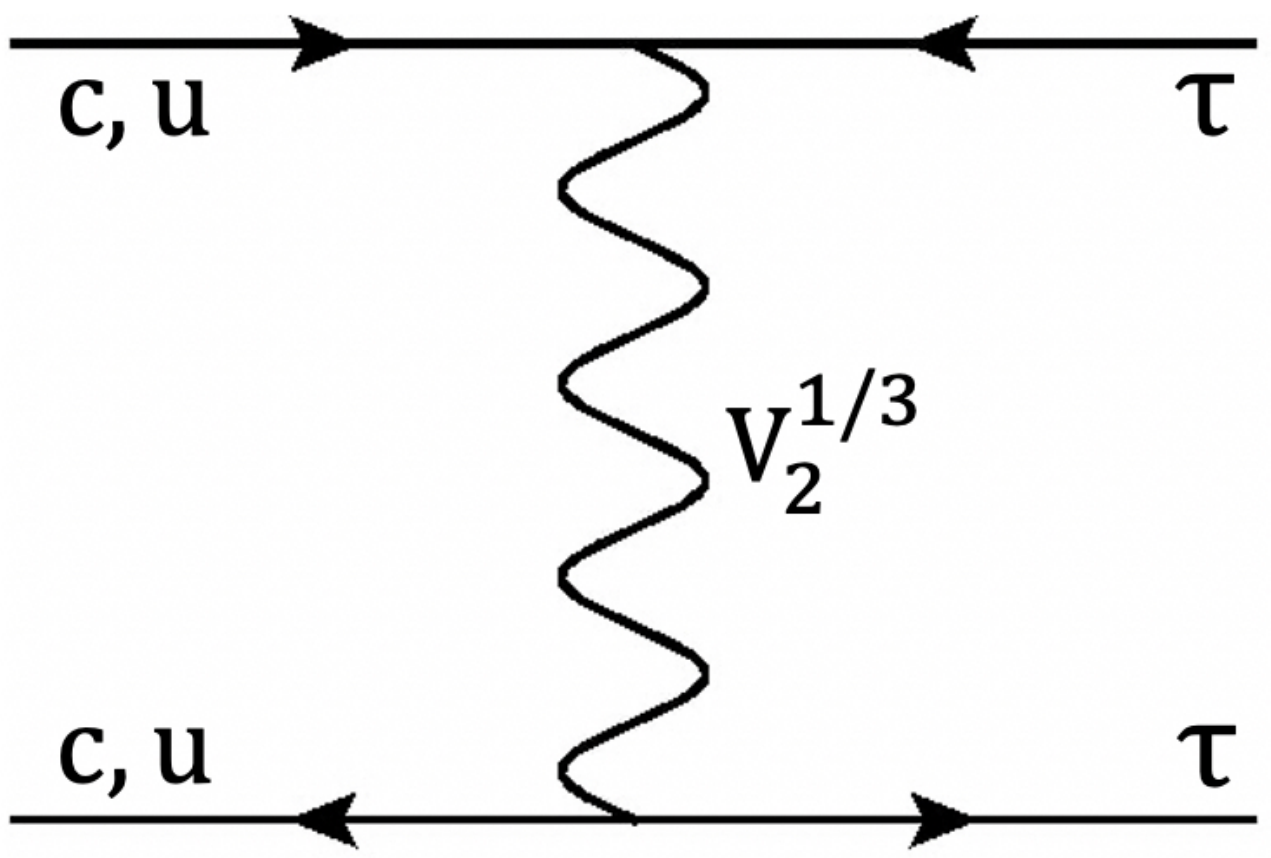}
\end{center}
 \caption{
  \label{fig:diagrams2}
  The contributing Feynman diagrams for $\tau\ov\tau$ final state at the LHC in the presence of non-vanishing $h_2^{13}$, $h_2^{23}$, $h_1^{33}$.
  In addition the one in Fig.\,\ref{fig:diagrams}, $h_2^{13}$ induces $d \ov d\to \tau \ov\tau$ and $u\ov u\to \tau\ov\tau$ processes mediated by $\rm{V}^{4/3}_2$ and $\rm{V}^{1/3}_2$.
 }
\end{figure}
In this section, we investigate $\mathcal{O} (\epsilon)$ contributions and derive the upper limit on the coupling products.
As is summarized in Tab.\,\ref{Tab:ModelNRC}, there are several processes to be discussed.

First, we consider $h_2^{13}$ contribution. If $h_2^{13}$ is sizable, the coupling contributes to $B_u\to\tau\ov\nu_\tau$.  
The contribution to $C_{S_R}^{ub}$ is expressed as 
\begin{align}
C_{S_R}^{ub}(\mu_{\rm{LQ}})= -\frac{1}{\sqrt{2}G_F V_Q^{cb}}\frac{h_1^{33}\left(V_Q^{11}h_2^{13*}+V_Q^{12}h_2^{23*}\right)}{m_{\rm{V}_2}^2}.
\end{align}
If $h_2^{13*}=-V_Q^{12}/V_Q^{11}h_2^{23*}\simeq -0.23h_2^{23*}$ is satisfied, $C_{S_R}^{ub}$ can be enough small to evade the bound from $R_{B_u}$ mentioned above and rescue the solution.

The tree-level exchange of the LQ induces $B_d\to \tau \ov\tau$, and the amplitude is proportional to $h_1^{33} \times h_2^{13*}$.
The $B_d\to \tau\ov\tau$ contribution has a chirality enhancement, so that it gives a strong bound.
The current experimental upper limit is BR($B_d\to\tau\ov\tau)\le 2.1 \times 10^{-3}$ at 95\,$\%$ CL. \cite{PDG2022} and the Belle II experiment is expected to probe BR($B_d\to\tau\ov\tau)\simeq 9.6 \times 10^{-5}$ \cite{Kou:2018nap}.
We note that $B\to\pi\tau\ov\nu_\tau$ is also induced by the same operator but the bound is weak. 
In addition, $c\ov u\tau\ov \tau$ and $s\ov d\tau\ov\tau$ four-fermi interactions are lead by the tree-level LQ exchange 
and the coefficients are proportional to $h_2^{13} \times h_2^{23*}$.
The couplings, however, do not predict rare meson decays since $\tau$ mass does not allow decay processes such as $D\to\tau\ov\tau$ nor $D\to \pi \tau\ov\tau$.\footnote{The one-loop contribution contributes to $K-\ov{K}$ mixing, but the abound is not so tight since there is no chirality enhancement as long as $h_1^{13}\times h_1^{23*}$ is small.
Furthermore for the correct loop calculation, we need the UV model and hence we limit ourselves to focus on the tree level phenomenology in this paper. 
We will come back to this point in Sec.\,\ref{sec:Summary}.}

Our numerical analysis shows that the $R_{D^{(*)}}$ anomaly can be explained within 2$\,\sigma$, if $h_1^{33}\times h_2^{23*}$ is fixed within $[-0.29,\,-0.12]$ when $m_{\rm{V}_2}=2\,$TeV.
Let us define the ratio of $h_2^{13}$ to $h_2^{23}$ as $h_2^{13}=-\lambda^{uc} h_2^{23}$.
This ratio is limited by $B_u\to\tau\ov\nu_\tau$.
When $h_1^{33}\times h_2^{23*}$ is around $-0.29$ ($-0.12$),
$\lambda^{uc}$ should satisfy $0.07\lesssim\lambda^{uc}\lesssim 0.57$ ($0.16\lesssim\lambda^{uc}\lesssim 0.37$) to evade the $B_u\to\tau\ov\nu_\tau$ bound.
In the future, the Belle II experiment could improve the bound and this range could be reduced to be $0.20\lesssim\lambda^{uc}\lesssim 0.34$ ($0.16\lesssim\lambda^{uc}\lesssim 0.48$) if the experimental central value does not change \cite{Kou:2018nap}.

\begin{figure}[t]
\begin{center}
 \includegraphics[width=0.32 \textwidth]{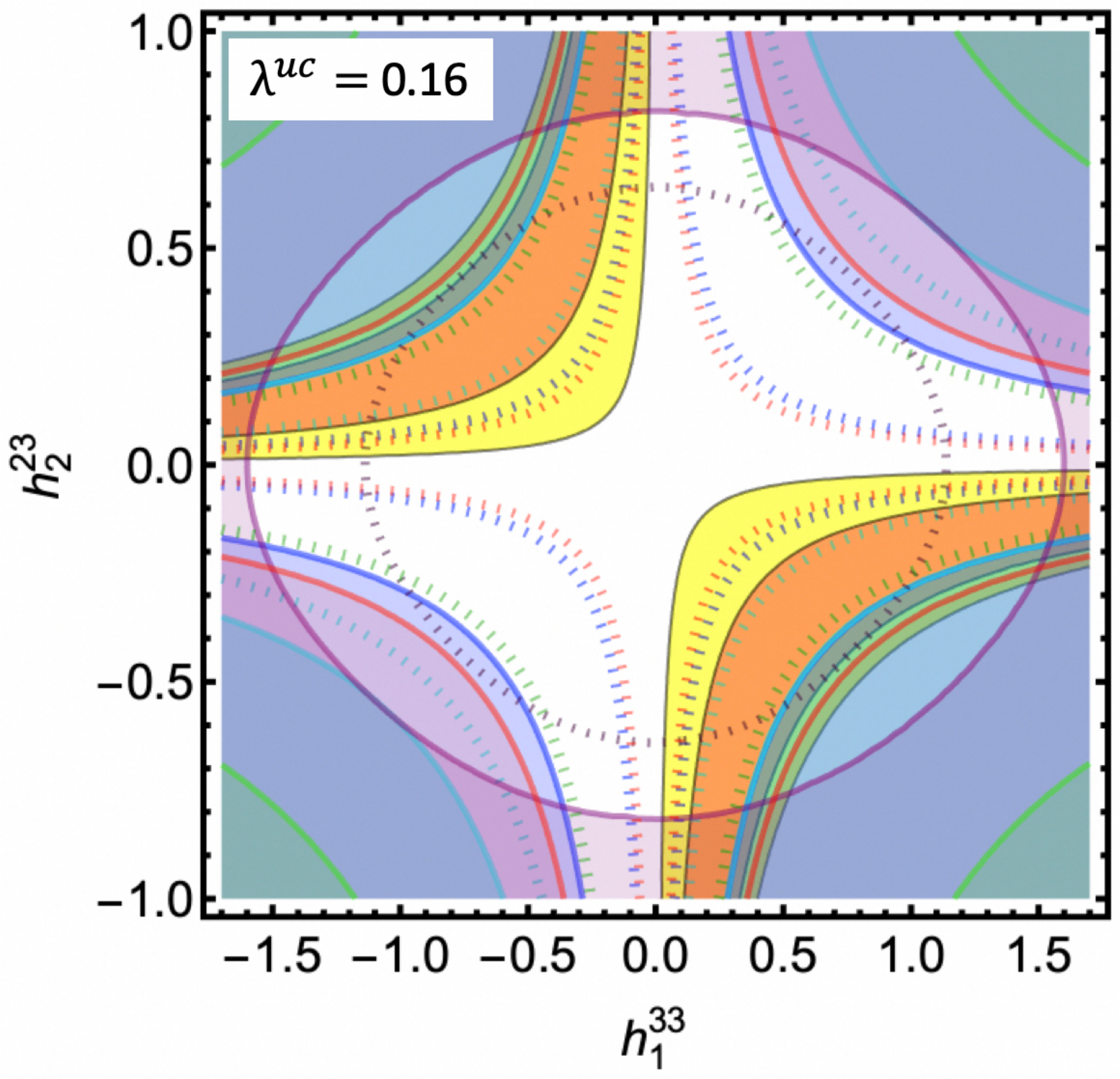}~~
 \includegraphics[width=0.32 \textwidth]{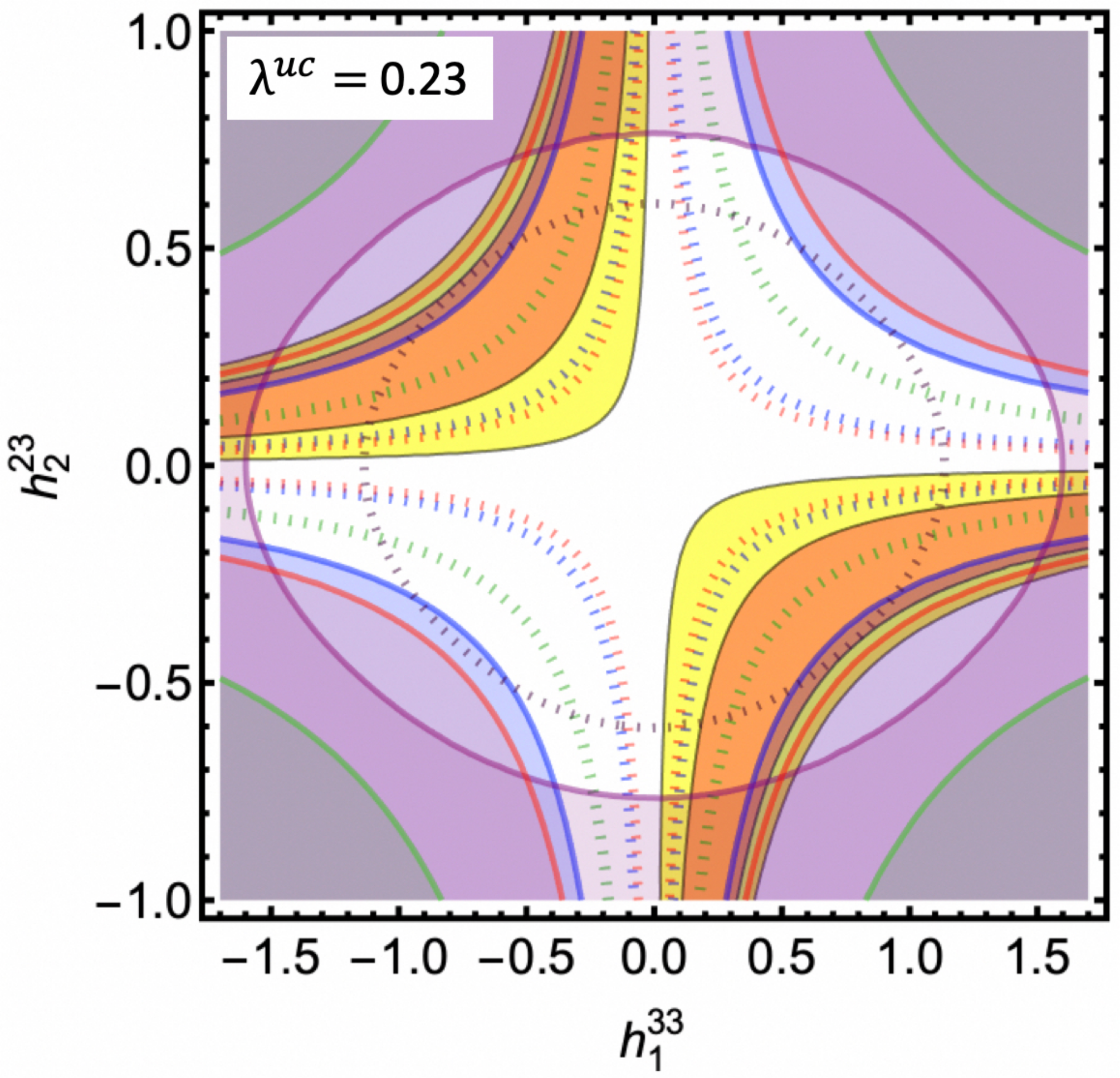}~~
 \includegraphics[width=0.32 \textwidth]{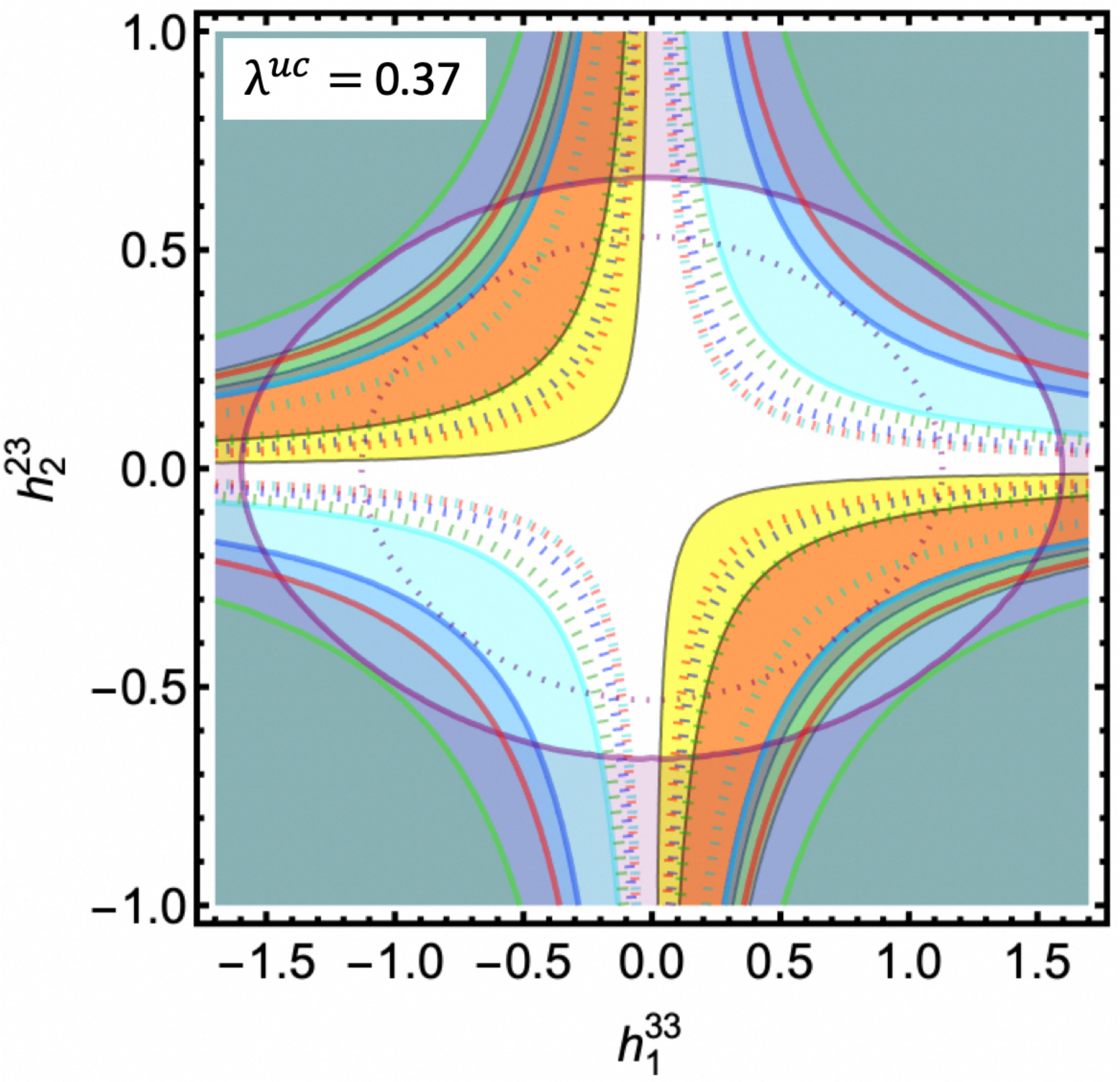}
\end{center}
 \caption{
  \label{fig:h133h223h213}
  The color code is the same as in Fig.\,\ref{fig:h133_h223}.
  Additionally, cyan and light green regions show the exclusion from $B_u\to\tau\ov\nu_\tau$ and $B_d\to\tau\ov\tau$.
  Their future prospects are also shown in dotted lines.
  }
\end{figure}
$h_2^{13}$ also contributes to $C_{S_R}^{cb}$ as
\begin{align}
C_{S_R}^{cb}(\mu_{\rm{LQ}})= -\frac{1}{\sqrt{2}G_F V_Q^{ub}}\frac{h_1^{33}\left(V_Q^{22}h_2^{23*}+V_Q^{21}h_2^{13*}\right)}{m_{\rm{V}_2}^2},
\end{align}
although it is suppressed by $V_Q^{21}$.
Besides, $h_2^{13}$ also contributes to $d\ov d\to \tau\ov\tau$ and $u\ov u\to \tau\ov\tau$ processes at the LHC as shown in Fig.\,\ref{fig:diagrams2}.
It is noted that $h_2^{33}$ contributes to $C_{S_R}^{cb}$ with the CKM factor, $V_Q^{23}$.
Due to the off-diagonal CKM suppression, both $h_1^{33}$ and $h_2^{33}$ should be sizable to enhance $C_{S_R}^{cb}$.
This additional entry does not affect $B_s\to \tau\ov\tau$ and $B\to K\tau\ov\tau$ while contributes to high-$p_T$ observables via $b\ov b \to \tau\ov\tau$.
As a result, the impact on $R_{D^{(*)}}$ is small compared to that from $h_2^{23}$.

\begin{figure}[t]
\begin{center}
 \includegraphics[width=0.32 \textwidth]{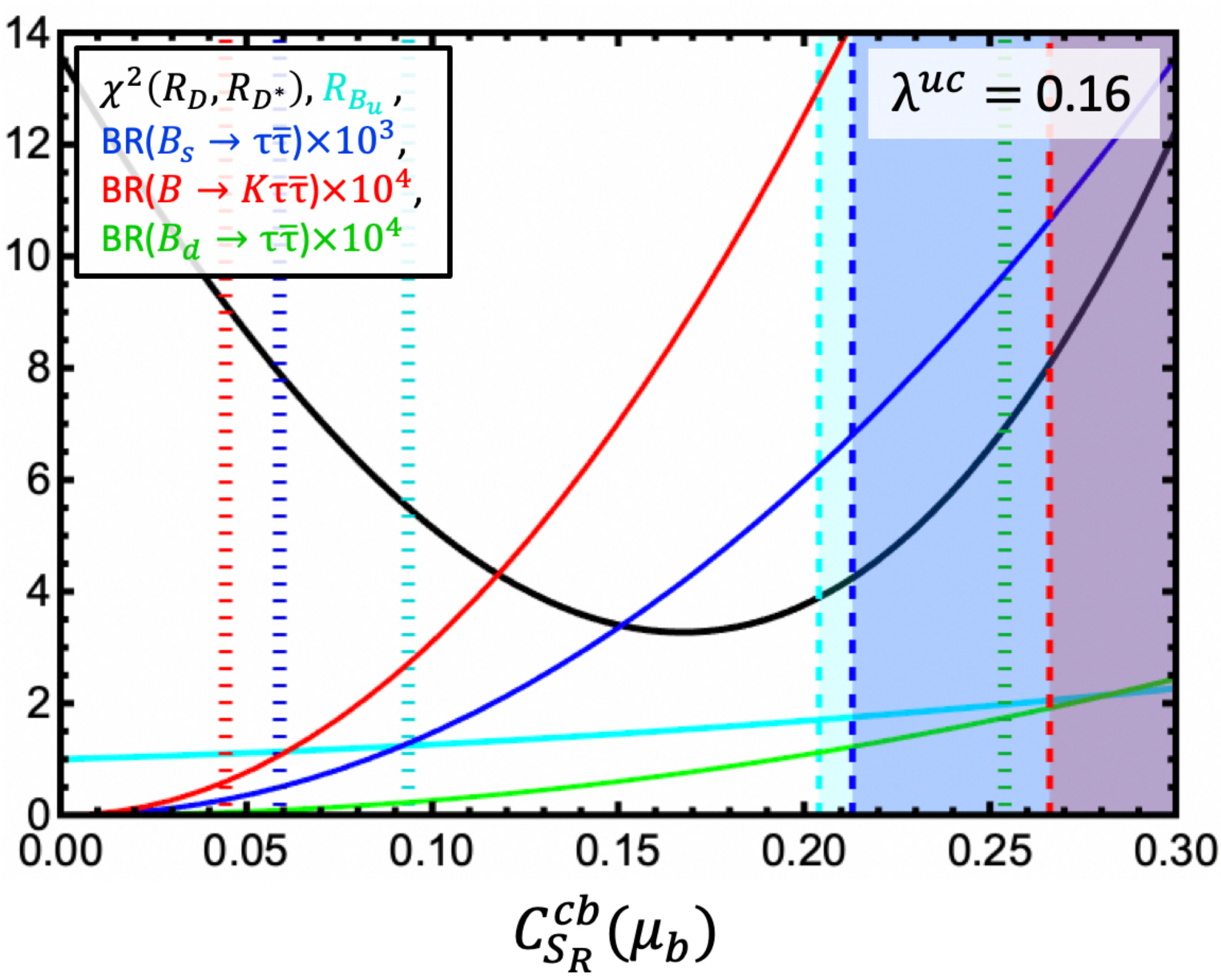}~~
 \includegraphics[width=0.32 \textwidth]{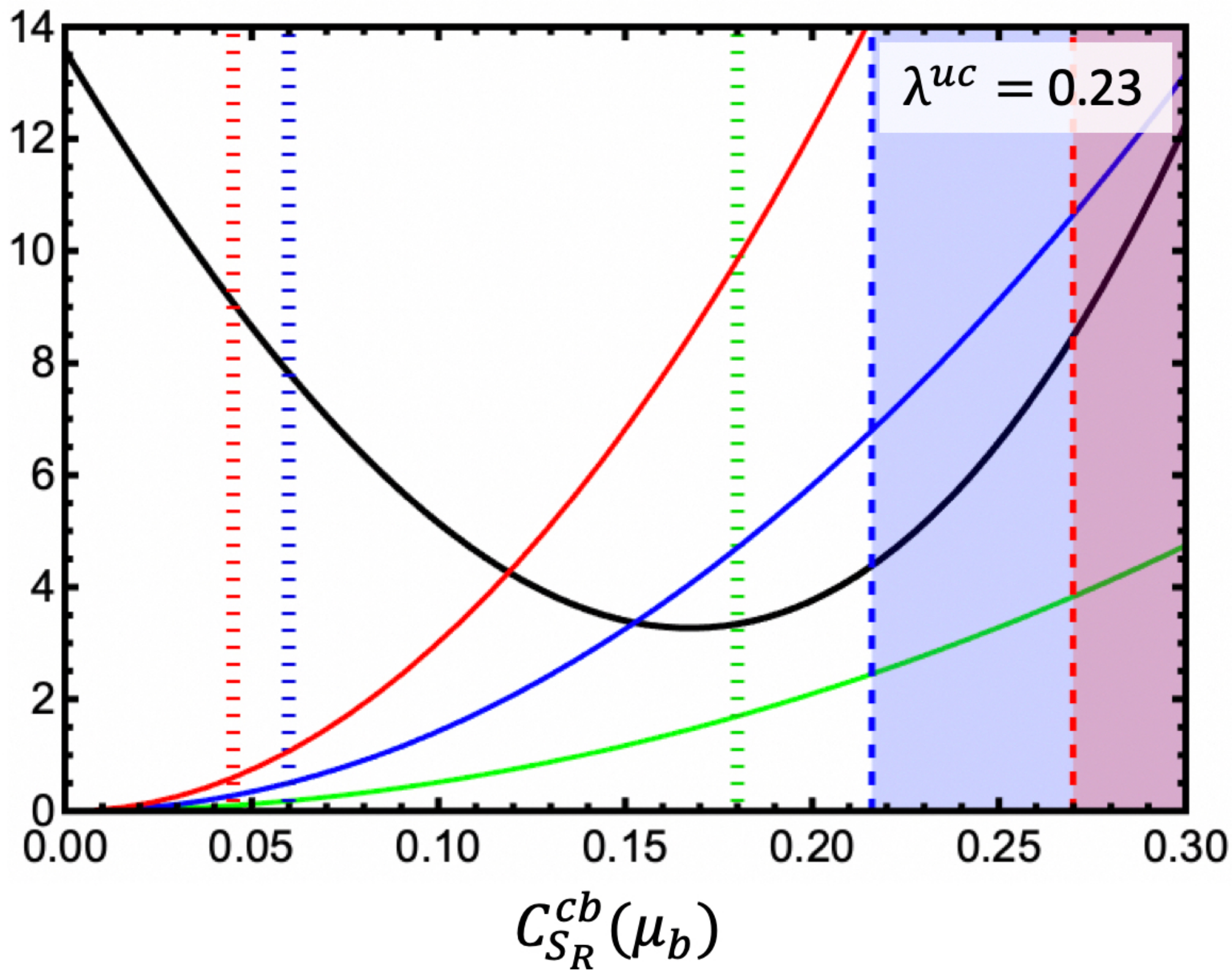}~~
 \includegraphics[width=0.32 \textwidth]{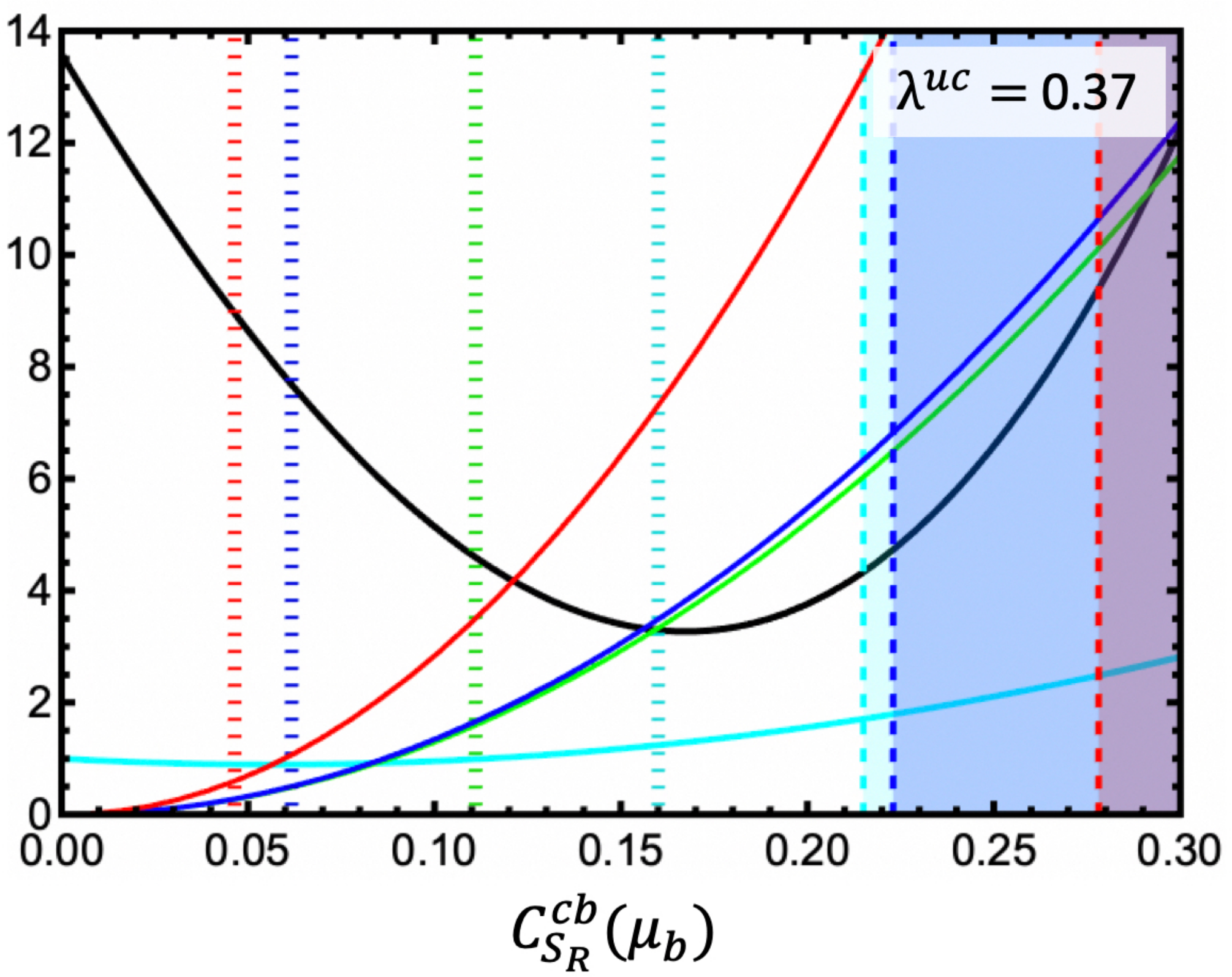}
\end{center}
 \caption{
  \label{fig:Obs_correlation}
The correlation of prediction is shown as a function of $C_{S_R}^{cb}(\mu_b)$ fixing $\lambda^{uc}=0.16$ (left), 0.23 (middle) and 0.37 (right).
  $\chi^2(R_D,\,R_{D^*})$, $R_{B_u}$ is shown in black and cyan.
  BR$(B_s\to\tau\ov\tau)\times 10^3$, BR$(B\to K\tau\ov\tau)\times 10^4$ and BR$(B_d\to\tau\ov\tau)\times 10^4$ are shown in blue, red and light green lines.
  The current exclusion is shown in each color while the future prospect is shown in dotted line.
  }
\end{figure}

In Fig.\,\ref{fig:h133h223h213}, $\lambda^{uc}$ is fixed at $\lambda^{uc}=0.16,\,0.23,\,0.37$ on the left, middle and right panels, respectively.
To see the prediction to motivate future experiments, the correlation among $\chi^2(R_D, R_{D^*})$, $R_{B_u}$, BR$(B_s\to\tau\ov\tau)\times 10^3$, BR$(B\to K\tau\ov\tau)\times 10^4$ and BR$(B_d\to\tau\ov\tau)\times 10^4$ is shown in Fig.\,\ref{fig:Obs_correlation}.
We see that in addition to $B_s\to\tau\ov\tau$ and $B\to K\tau\ov\tau$, $B_d\to\tau\ov\tau$ plays an important role in the probe when $\lambda^{uc}=0.23$ and 0.37.
When $\lambda^{uc}=0.16$ and 0.37, the future $R_{B_u}$ measurement can probe the best fit point of the model.
On the other hand, $R_{B_u}$ is suppressed when $\lambda^{uc}=0.23$, because of the cancellation mentioned above.
We note that that real $h_1^{33}\times h_2^{23*}$ is favored by the current $R_{D^{(*)}}$ measurement.
If $h_1^{33}\times h_2^{13*}$ is also real, we obtain the prediction for $R_{B_u}$ as $0.89\lesssim R_{B_u}$, due to the relative phase between the SM amplitude and $\rm{V}_2$ amplitude.
We note that the collider reach is also mildly extended with the inclusion of non-vanishing $h_2^{13}$.

We investigate the phenomenological impact of other couplings.
As shown in Tab.\,\ref{Tab:ModelNRC}, sizable $h_1^{13}$ and $h_2^{23*}$ couplings
predict the contributions to the leptonic $D$ meson decays: $D_{d,s}\to\tau\ov\nu_\tau$, corresponding to the category (iv).
To study the bounds numerically, we define
\begin{align}
R_{D_d}=\frac{{\rm{BR}}(D_d\to \tau\ov\nu_\tau)}{{\rm{BR}}(D_d\to \tau\ov\nu_\tau)_{\rm SM}} =\left| 1 + 1.47 C^{cd}_{S_R} \right|^2 ,
\label{eq:D}
\end{align}
and
\begin{align}
 \label{eq:Ds}
R_{D_s}=\frac{{\rm{BR}}(D_s\to \tau\ov\nu_\tau)} {{\rm{BR}}(D_s\to \tau\ov\nu_\tau)_{\rm SM}}=\left| 1 + 1.72  C_{S_R}^{cs} \right|^2,
\end{align}
where BR$(D_d\to \tau\ov\nu_\tau)_{\rm SM}\simeq1.06 \times 10^{-3}$ and BR$(D_s\to \tau\ov\nu_\tau)_{\rm SM} \simeq 5.2 \times 10^{-2}$. The numerical estimations are obtained using $|V_{cd}|=0.221\pm0.004$, $|V_{cs}|=0.975\pm0.006$ \cite{PDG2022}, $f_{D}=0.2120\pm0.007$\,GeV and  $f_{D_s}=0.2499\pm0.005$\,GeV \cite{HFLAV:2022pwe}.
Compared to $B_u$ and $B_c$ decays, those enhancement factors are small where $m_c(m_c)=1.27$\,GeV is used. $C_{S_R}^{cs}$ is given by Eq. (\ref{eq:CSR_general}),
and proportional to $h_1^{13}\times h_2^{23*}$.

There are experimental results of those decays: BR$(D_d\to \tau\ov\nu_\tau)_{\rm exp}=(1.20\pm0.27)\times 10^{-3}$ and  BR$(D_s\to \tau\ov\nu_\tau)_{\rm exp}=5.32\pm0.11\,\%$ \cite{PDG2022}.
Furthermore, the BES III experiment reports the recent result \cite{BESIII:2019vhn}: BR$(D_d\to\tau\ov\nu_\tau)$/BR$(D_d\to\mu\ov\nu_\mu)=3.21(1\pm0.24)$ where the SM prediction is 2.67. 
When we require $C_{S_R}^{cs}$ to satisfy $0.5\le {\rm{BR}}(D_d\to \tau\ov\nu_\tau)/{\rm{BR}}(D_d\to \tau\nu_\tau)_{\rm{SM}}\le1.5$ and $0.95\le{\rm{BR}}(D_s\to \tau\ov\nu_\tau)/{\rm{BR}}(D_s\to \tau\nu_\tau)_{\rm{SM}}\le1.05$, we obtain 
\begin{align}
   -1.3 \lesssim h_1^{13}\times h_2^{23*}\lesssim 1.1,~~~ 
   -0.46 \lesssim h_1^{23}\times h_2^{23*}\lesssim 0.48.
\end{align}

Sizable $h_1^{13}$ and $h_1^{23}$ enhances or suppresses $B\to K^{(*)} \nu_\tau\ov\nu_\tau$ and $B\to \pi \nu_\tau\ov\nu_\tau$ corresponding to the category 
(ii).
The contributions of the LQ exchange are proportional to $h_1^{33}\times h_1^{23*}$ and $h_1^{33}\times h_1^{13*}$, respectively.\footnote{We find that $B\to \pi \tau\ov\tau$ is less constraining compared to $B_s\to\tau\tau$ \cite{Capdevila:2017iqn}. In fact, the bound on $B\to \pi \tau\ov\tau$ is not available in PDG \cite{PDG2022}.
}
The LQ contributions correspond to the vector operators.
It is conventional to define the ratio as $\mathcal{R}_{M_1}^{\nu} = {\rm{BR}}(B\to M_1\nu\ov\nu )/{\rm{BR}}(B\to M_1 \nu\ov\nu )_{\rm{SM}}$.
The Belle collaboration has provided an upper bound as 
 $\mathcal{R}_{K^*}^{\nu} \le 2.7$ and $\mathcal{R}_{K}^{\nu } \le 3.9$ at the $90\%$ CL.~\cite{Belle:2017oht}.
The Belle II could measure the SM prediction with $10\,\%$ accuracy \cite{Kou:2018nap}.
Following Ref.\,\cite{Buras:2014fpa}, those ratios are expressed as 
\begin{align}
\mathcal{R}_{K^*}^{\nu} =(1-2\eta_{\nu})\epsilon_{\nu}^2,~~
\mathcal{R}_{K}^{\nu} =(1+\kappa_{\nu}\eta_{\nu\nu})\epsilon_{\nu}^2,
\end{align}
where
\begin{align}
\epsilon_{\nu}=\frac{\sqrt{|C_L^{\rm SM}|^2+|C_{R}^{bs}|^2}}{|C_L^{\rm SM}|},~~
\eta_{\nu}=\frac{-\rm{Re}\left( C_L^{\rm SM}C_{R}^{bs*} \right)}{|C_L^{\rm SM}|^2+|C_{R}^{bs}|^2},
\end{align}
and $\kappa_{\nu}= 1.34 \pm 0.04$ with $C_L^{\rm SM}\simeq-6.35\,$.
Similarly the current upper limit on $B\to \pi\nu\ov \nu$ is approximately given as 
\begin{align}
\mathcal{R}_{\pi}^{\nu}\lesssim 6000,
\end{align}
using the SM prediction of ${\rm{BR}}(B^0\to\pi^0 \nu\ov \nu)_{{\rm{SM}}}=(5.4\pm0.6)\times 10^{-8}$ \cite{Bause:2022rrs} and the current experimental constraint, ${\rm{BR}}(B^0\to\pi^0 \nu\ov \nu)\le0.9\times 10^{-5}$ at $90\,\%$ C.L. \cite{PDG2022}.
Those upper bounds lead to 
\begin{align}
   -0.27 \lesssim h_1^{33}\times h_1^{23*}\lesssim 0.21,~~
   -3.7 \lesssim h_1^{33}\times h_1^{13*}\lesssim 3.7.
\end{align}
We find that the constraints from $B_d\to\tau\ov\tau$ and $B_s\to\tau\ov\tau$ lead to $-1.8 \lesssim h_1^{33}\times h_1^{13*} \lesssim 1.8$ and $-2.65 \lesssim h_2^{33}\times h_2^{23*} \lesssim 2.65$, respectively.

The bounds on the coupling products are summarized 
in Tab.\,\ref{Tab:ModelNRC_summary}.
We see that there are several combinations that are allowed to be $\mathcal{O}(1)$.
Especially, it is not easy to constrain $h_1^{33}\times h_1^{13*}$, $h_2^{13}\times h_2^{23*}$ and $h_2^{33}\times h_2^{23*}$ using tree level processes.
In other words $h_{2}^{33}$ and $h_1^{13}$ can be of order 1 while $h_2^{13}$ and $h_1^{23}$ should be somewhat smaller.
We may obtain strong bounds considering one-loop contributions.
Such a higher-order contribution usually involves extra fields in the loop, so a concrete setup needs to be taken into account.

Finally, we discuss the bound from the collider experiments.
When $h_1^{13}$ and $h_2^{13}$, that correspond to the couplings involving light quarks, are sizable, our model 
can be constrained by the collider searches further.
We study the bounds from the di-$\tau$ and mono-$\tau$ signatures at the LHC by repeating the procedure explained in Sec.\,\ref{sec:LHC}.
As shown in Figs.\,\ref{fig:diagrams2}, the t-channel diagrams given by the exchange of the LQ induce di-$\tau$ signatures.
Analogously we have t-channel diagrams contributing to mono-$\tau$ signature.
Based on Run\,2 full data  we derive the upper limit on the coupling at $2\sigma$ as follows :
\begin{eqnarray}
|h_1^{13}|\le0.55,&& |h_2^{13}|\le0.51,~ |h_1^{23}|\le1.01,~ |h_2^{23}|\le0.93, \nonumber \\ |h_1^{33}|\le1.60,&& |h_2^{33}|\le1.66.
\end{eqnarray}
In this analysis, we turn on the only one coupling assuming that the other couplings are vanishing.
We see that $h_{1,2}^{13}$ coupling can be at most 0.5.
Also we can set the upper limit on the least constrained $h_2^{33}$ with LHC data.
This collider constraint is complementary to the flavor constraint.
We introduce a Yukawa texture, that respects the $\tau$ number, as an illustration:\footnote{We note that $h^{33}_{2}$ could be sizable as shown in Table \ref{Tab:ModelNRC_summary}.}
\begin{eqnarray}
  h_1^{ij} \simeq\left(
  \begin{array}{ccc}
   ~0~&~0~&~\mathcal{O}(10^{-3})\\
   ~0~&~0~&~\mathcal{O}(10^{-2})\\
   ~0~&~0~&~\mathcal{O}(1)\\
  \end{array}
  \right),~~~h_2^{ij} \simeq\left(
  \begin{array}{ccc}
   ~0~&~0~&~-0.23 h_2^{23}\\
   ~0~&~0~&~\mathcal{O}(1)\\
   ~0~&~0~&~\mathcal{O}(0.1) \\
  \end{array}
  \right).
\label{eq:coupling_structure_2023}
\end{eqnarray}
\begin{table}[t]
\begin{center}
\scalebox{0.95}{
  \begin{tabular}{c|c} \hline
   Coupling product & bound \\ \hline \hline 
 $h_1^{33}\times h_2^{23*}$ &   [$-0.28,~-0.12$]\\ \hline\hline
  $h_1^{33}\times h_2^{13*}$ &  [$-0.02,~0.04$] \\  \hline
  $h_1^{33}\times h_2^{33*}$ &  --- \\ \hline
  $h_1^{33}\times h_1^{13*}$ &  [$-1.8,~1.8$] \\ \hline
  $h_1^{33}\times h_1^{23*}$ &  [$-0.27,~0.21$] \\ \hline
  $h_1^{13}\times h_2^{23*}$ & [$-1.3,~1.1$] \\ \hline
  $h_1^{23}\times h_2^{23*}$ &  [$-0.46,~0.48$] \\ \hline
  $h_2^{13}\times h_2^{23*}$ &  --- \\ \hline
  $h_2^{33}\times h_2^{23*}$ &  [$-2.65,~2.65$] \\ \hline
   \end{tabular}
  }
  \caption{Summary table for the non-LHC bound on the coupling product assuming $m_{\rm{V}_2}=2\,$TeV.
}
  \label{Tab:ModelNRC_summary}
\end{center}   
\vspace{-.45cm}
\end{table}

\section{Summary and discussion}
\label{sec:Summary}
In this paper we studied phenomenology of the model with isodoublet vector LQ, $\rm{V}_2$.
In light of the recent result of $R_{D^{(*)}}$, the LQ becomes very interesting.
$\chi^2(R_D,\,R_{D^*})$ can be as small as 3.7 in this model and the minimal coupling scenario predicts that $B_s\to\tau\ov\tau$ and $B\to K\tau\ov\tau$ within the reach of the Run\,3 LHCb measurement and early Belle II with 5\,ab$^{-1}$, respectively. 
In the minimal setup, $B_u\to\tau\ov\nu_\tau$ is deviated from the SM prediction, so that the setup is excluded.
This bound can be evaded by introducing another flavor violating coupling to the large contribution to $B_u\to\tau\ov\nu_\tau$.
We conclude that there are setups that are consistent with the experimental results related to the flavor physics as well as the high-$p_T$ signals.

We only discussed the tree-level contributions induced by the $\rm{V}_2$ exchange.
The LQ mass is not large, so it may be necessary to take into account the one-loop corrections involving $\rm{V}_2$. The study, however, would requires a complete model since the loop diagrams involve extra fields e.g. extra fermions and scalars in general and the contributions would not be negligible \cite{Iguro:2022ozl}. 
It would be challenging to construct a complete model with $\rm{V}_2$, since the constraint from the lifetime of proton is very strong and a specific parameter setup is required to explain the $R_{D^{(*)}}$ anomaly.
The quantum number of $\rm{V}_2$ is the same as $X$ boson in the SU(5) GUT.
We could, for instance, consider the model where the SU(5) unification is realized in only one generation: the fields in the other generations are charged under G$^\prime_{\rm SM}=$SU(3)$_c\times$SU(2)$_L\times$U(1)$_Y$.
The SM gauge symmetry is given by the linear combination of G$^\prime_{\rm SM}$ and the subgroups of SU(5).
If SU(5)$\times$G$^\prime_{\rm SM}$ breaks down to the SM gauge symmetry at the low scale, 
$\rm{V}_2$ would arise as a massive gauge boson with a light mass.
In this setup, $\rm{V}_2$ could approximately have a quantum number like the $\tau$ number.
The couplings of $\rm{V}_2$ with light fermions may be suppressed and may be able to suppress the dangerous couplings that cause proton decay, at the tree level. The fields and couplings to realize the realistic fermion mass matrices, however, may cause additional contributions to flavor physics at the tree and the one-loop levels, as discussed in Ref. \cite{Iguro:2022ozl}. The constraints from searches for the particles predicted by the underlying theory may disturb the $R_{D^{(*)}}$ anomaly explanation \cite{Iguro:2021kdw}. 
We need further detailed study \cite{IOprojectNo17}.

\section*{Acknowledgements}
The authors would like to thank Ulrich Nierste and Teppei Kitahara for the discussion.
We also appreciate Felix Wilsch for the great supports on HighPT.
S.I. enjoys the support from the Deutsche Forschungsgemeinschaft (DFG, German Research Foundation) under grant 396021762-TRR\,257.
The work of Y.\,O.~is supported by Grant-in-Aid for Scientific research from the MEXT, Japan, No.\,19K03867.

\small{
\bibliographystyle{utphys28mod}
\bibliography{ref}

\providecommand{\href}[2]{#2}\begingroup\raggedright\begin{thebibliography}{100}

\bibitem{Bordone:2019guc}
M.~Bordone, N.~Gubernari, D.~van Dyk, and M.~Jung, ``{Heavy-Quark expansion for
  ${{\bar{B}}_s\rightarrow D^{(*)}_s}$ form factors and unitarity bounds beyond
  the ${SU(3)_F}$ limit},''
  \href{https://dx.doi.org/10.1140/epjc/s10052-020-7850-9}{Eur.\  Phys.\  J.\
  C {\bfseries 80} (2020) 347} {\ttfamily
  [\href{https://arxiv.org/abs/1912.09335}{arXiv:1912.09335}]}.

\bibitem{Iguro:2020cpg}
S.~Iguro and R.~Watanabe, ``{Bayesian fit analysis to full distribution data of
  $ \overline{\mathrm{B}}\to {\mathrm{D}}^{\left(\ast
  \right)}\mathrm{\ell}\overline{\nu }:\left|{\mathrm{V}}_{\mathrm{cb}}\right|
  $ determination and new physics constraints},''
  \href{https://dx.doi.org/10.1007/JHEP08(2020)006}{JHEP {\bfseries 08} (2020)
  006} {\ttfamily [\href{https://arxiv.org/abs/2004.10208}{arXiv:2004.10208}]}.

\bibitem{HFLAV:2022pwe}
{\bfseries HFLAV} Collaboration, ``{Averages of $b$-hadron, $c$-hadron, and
  $\tau$-lepton properties as of 2021}.'' {\ttfamily
  \href{https://arxiv.org/abs/2206.07501}{arXiv:2206.07501}}. {Average of $R_D$
  and $R_{D^\ast}$ for Winter 2023 at
  \url{https://hflav-eos.web.cern.ch/hflav-eos/semi/winter23_prel/html/RDsDsstar/RDRDs.html}}.

\bibitem{Bernlochner:2022ywh}
F.~U.~Bernlochner, { et al.}, ``{Constrained second-order power corrections in
  HQET: R(D(*)), $|V_{cb}|$, and new physics},''
  \href{https://dx.doi.org/10.1103/PhysRevD.106.096015}{Phys.\  Rev.\  D
  {\bfseries 106} (2022) 096015} {\ttfamily
  [\href{https://arxiv.org/abs/2206.11281}{arXiv:2206.11281}]}.

\bibitem{Martinelli:2021onb}
G.~Martinelli, S.~Simula, and L.~Vittorio, ``{$\vert V_{cb} \vert$ and
  $R(D)^{(*)}$) using lattice QCD and unitarity},''
  \href{https://dx.doi.org/10.1103/PhysRevD.105.034503}{Phys.\  Rev.\  D
  {\bfseries 105} (2022) 034503} {\ttfamily
  [\href{https://arxiv.org/abs/2105.08674}{arXiv:2105.08674}]}.

\bibitem{Martinelli:2021myh}
G.~Martinelli, S.~Simula, and L.~Vittorio, ``{Exclusive determinations of
  $\vert V_{cb} \vert $ and $R(D^{*})$ through unitarity},''
  \href{https://dx.doi.org/10.1140/epjc/s10052-022-11050-0}{Eur.\  Phys.\  J.\
  C {\bfseries 82} (2022) 1083} {\ttfamily
  [\href{https://arxiv.org/abs/2109.15248}{arXiv:2109.15248}]}.

\bibitem{FermilabLattice:2021cdg}
{\bfseries Fermilab Lattice, MILC, Fermilab Lattice, MILC} Collaboration,
  ``{Semileptonic form factors for $B\rightarrow D^*\ell \nu $ at nonzero
  recoil from $2+1$-flavor lattice QCD: Fermilab
  Lattice~and~MILC~Collaborations},''
  \href{https://dx.doi.org/10.1140/epjc/s10052-022-10984-9}{Eur.\  Phys.\  J.\
  C {\bfseries 82} (2022) 1141} {\ttfamily
  [\href{https://arxiv.org/abs/2105.14019}{arXiv:2105.14019}]}. [Erratum:
  Eur.Phys.J.C 83, 21 (2023)].

\bibitem{Fedele:2023ewe}
M.~Fedele, { et al.}, ``{Discriminating $B\to D^{*}\ell\nu$ form factors via
  polarization observables and asymmetries}.'' {\ttfamily
  \href{https://arxiv.org/abs/2305.15457}{arXiv:2305.15457}}.

\bibitem{Lees:2012xj}
{\bfseries BaBar} Collaboration, ``{Evidence for an excess of $\bar{B} \to
  D^{(*)} \tau^-\bar{\nu}_\tau$ decays},''
  \href{https://dx.doi.org/10.1103/PhysRevLett.109.101802}{Phys.\  Rev.\
  Lett.\  {\bfseries 109} (2012) 101802} {\ttfamily
  [\href{https://arxiv.org/abs/1205.5442}{arXiv:1205.5442}]}.

\bibitem{Lees:2013uzd}
{\bfseries BaBar} Collaboration, ``{Measurement of an Excess of $\bar{B} \to
  D^{(*)}\tau^- \bar{\nu}_\tau$ Decays and Implications for Charged Higgs
  Bosons},'' \href{https://dx.doi.org/10.1103/PhysRevD.88.072012}{Phys.\  Rev.\
   D {\bfseries 88} (2013) 072012} {\ttfamily
  [\href{https://arxiv.org/abs/1303.0571}{arXiv:1303.0571}]}.

\bibitem{Huschle:2015rga}
{\bfseries Belle} Collaboration, ``{Measurement of the branching ratio of
  $\bar{B} \to D^{(\ast)} \tau^- \bar{\nu}_\tau$ relative to $\bar{B} \to
  D^{(\ast)} \ell^- \bar{\nu}_\ell$ decays with hadronic tagging at Belle},''
  \href{https://dx.doi.org/10.1103/PhysRevD.92.072014}{Phys.\  Rev.\  D
  {\bfseries 92} (2015) 072014} {\ttfamily
  [\href{https://arxiv.org/abs/1507.03233}{arXiv:1507.03233}]}.

\bibitem{Hirose:2016wfn}
{\bfseries Belle} Collaboration, ``{Measurement of the $\tau$ lepton
  polarization and $R(D^*)$ in the decay $\bar{B} \to D^* \tau^-
  \bar{\nu}_\tau$},''
  \href{https://dx.doi.org/10.1103/PhysRevLett.118.211801}{Phys.\  Rev.\
  Lett.\  {\bfseries 118} (2017) 211801} {\ttfamily
  [\href{https://arxiv.org/abs/1612.00529}{arXiv:1612.00529}]}.

\bibitem{Hirose:2017dxl}
{\bfseries Belle} Collaboration, ``{Measurement of the $\tau$ lepton
  polarization and $R(D^*)$ in the decay $\bar{B} \rightarrow D^* \tau^-
  \bar{\nu}_\tau$ with one-prong hadronic $\tau$ decays at Belle},''
  \href{https://dx.doi.org/10.1103/PhysRevD.97.012004}{Phys.\  Rev.\  D
  {\bfseries 97} (2018) 012004} {\ttfamily
  [\href{https://arxiv.org/abs/1709.00129}{arXiv:1709.00129}]}.

\bibitem{Abdesselam:2019dgh}
{\bfseries Belle} Collaboration, ``{Measurement of $\mathcal{R}(D)$ and
  $\mathcal{R}(D^{\ast})$ with a semileptonic tagging method}.'' {\ttfamily
  \href{https://arxiv.org/abs/1904.08794}{arXiv:1904.08794}}.

\bibitem{Belle:2019rba}
{\bfseries Belle} Collaboration, ``{Measurement of $\mathcal{R}(D)$ and
  $\mathcal{R}(D^*)$ with a semileptonic tagging method},''
  \href{https://dx.doi.org/10.1103/PhysRevLett.124.161803}{Phys.\  Rev.\
  Lett.\  {\bfseries 124} (2020) 161803} {\ttfamily
  [\href{https://arxiv.org/abs/1910.05864}{arXiv:1910.05864}]}.

\bibitem{Aaij:2015yra}
{\bfseries LHCb} Collaboration, ``{Measurement of the ratio of branching
  fractions $\mathcal{B}(\bar{B}^0 \to
  D^{*+}\tau^{-}\bar{\nu}_{\tau})/\mathcal{B}(\bar{B}^0 \to
  D^{*+}\mu^{-}\bar{\nu}_{\mu})$},''
  \href{https://dx.doi.org/10.1103/PhysRevLett.115.111803}{Phys.\  Rev.\
  Lett.\  {\bfseries 115} (2015) 111803} {\ttfamily
  [\href{https://arxiv.org/abs/1506.08614}{arXiv:1506.08614}]}. [Erratum:
  Phys.Rev.Lett. 115, 159901 (2015)].

\bibitem{Aaij:2017uff}
{\bfseries LHCb} Collaboration, ``{Measurement of the ratio of the $B^0 \to
  D^{*-} \tau^+ \nu_{\tau}$ and $B^0 \to D^{*-} \mu^+ \nu_{\mu}$ branching
  fractions using three-prong $\tau$-lepton decays},''
  \href{https://dx.doi.org/10.1103/PhysRevLett.120.171802}{Phys.\  Rev.\
  Lett.\  {\bfseries 120} (2018) 171802} {\ttfamily
  [\href{https://arxiv.org/abs/1708.08856}{arXiv:1708.08856}]}.

\bibitem{Aaij:2017deq}
{\bfseries LHCb} Collaboration, ``{Test of Lepton Flavor Universality by the
  measurement of the $B^0 \to D^{*-} \tau^+ \nu_{\tau}$ branching fraction
  using three-prong $\tau$ decays},''
  \href{https://dx.doi.org/10.1103/PhysRevD.97.072013}{Phys.\  Rev.\  D
  {\bfseries 97} (2018) 072013} {\ttfamily
  [\href{https://arxiv.org/abs/1711.02505}{arXiv:1711.02505}]}.

\bibitem{LHCb:2023zxo}
{\bfseries LHCb} Collaboration, ``{Measurement of the ratios of branching
  fractions $\mathcal{R}(D^{*})$ and $\mathcal{R}(D^{0})$}.'' {\ttfamily
  \href{https://arxiv.org/abs/2302.02886}{arXiv:2302.02886}}.

\bibitem{LHCbRun1had}
{\bfseries LHCb} Collaboration, ``{Rare and semileptinic decay $\&$ LFNU at
  LHCb}.''. {\url{https://agenda.infn.it/event/33694/contributions/190241/}}.

\bibitem{Ciezarek:2837207}
{\bfseries LHCb} Collaboration, ``{$R(D^\ast)$ and $R(D)$ with $\tau^- \to
  \mu^- \nu_\tau \overline\nu_\mu$}.''.
  {\url{https://indico.cern.ch/event/1187939/}}.

\bibitem{LHCbRDst2023}
R.~Puthumanaillam, ``{Measurement of $R(D^\ast)$ with hadronic $\tau^+$ decays
  at $\sqrt{s}=13$ TeV by the LHCb collaboration}.'' {Presentation at the CERN
  Seminar, CERN, 21 March 2023}.

\bibitem{Iguro:2022yzr}
S.~Iguro, T.~Kitahara, and R.~Watanabe, ``{Global fit to $b \to c\tau\nu$
  anomalies 2022 mid-autumn}.'' {\ttfamily
  \href{https://arxiv.org/abs/2210.10751}{arXiv:2210.10751}}.

\bibitem{Aebischer:2022oqe}
J.~Aebischer, G.~Isidori, M.~Pesut, B.~A.~Stefanek, and F.~Wilsch,
  ``{Confronting the vector leptoquark hypothesis with new low- and high-energy
  data},'' \href{https://dx.doi.org/10.1140/epjc/s10052-023-11304-5}{Eur.\
  Phys.\  J.\  C {\bfseries 83} (2023) 153} {\ttfamily
  [\href{https://arxiv.org/abs/2210.13422}{arXiv:2210.13422}]}.

\bibitem{Fedele:2022iib}
M.~Fedele, { et al.}, ``{Impact of
  \ensuremath{\Lambda}b\textrightarrow{}\ensuremath{\Lambda}c\ensuremath{\tau}\ensuremath{\nu}
  measurement on new physics in b\textrightarrow{}cl\ensuremath{\nu}
  transitions},'' \href{https://dx.doi.org/10.1103/PhysRevD.107.055005}{Phys.\
  Rev.\  D {\bfseries 107} (2023) 055005} {\ttfamily
  [\href{https://arxiv.org/abs/2211.14172}{arXiv:2211.14172}]}.

\bibitem{Aban:2023pgq}
J.~C.~Aban, C.-R.~Chen, and C.~S.~Nugroho, ``{Effects of Kaluza-Klein Neutrinos
  on $R_{D}$ and $R_{D^{*}}$}.'' {\ttfamily
  \href{https://arxiv.org/abs/2305.10305}{arXiv:2305.10305}}.

\bibitem{Beneke:1996xe}
M.~Beneke and G.~Buchalla, ``{The $B_c$ Meson Lifetime},''
  \href{https://dx.doi.org/10.1103/PhysRevD.53.4991}{Phys.\  Rev.\  D
  {\bfseries 53} (1996) 4991--5000} {\ttfamily
  [\href{https://arxiv.org/abs/hep-ph/9601249}{hep-ph/9601249}]}.

\bibitem{Alonso:2016oyd}
R.~Alonso, B.~Grinstein, and J.~Martin~Camalich, ``{Lifetime of $B_c^-$
  Constrains Explanations for Anomalies in $B\to D^{(*)}\tau\nu$},''
  \href{https://dx.doi.org/10.1103/PhysRevLett.118.081802}{Phys.\  Rev.\
  Lett.\  {\bfseries 118} (2017) 081802} {\ttfamily
  [\href{https://arxiv.org/abs/1611.06676}{arXiv:1611.06676}]}.

\bibitem{Celis:2016azn}
A.~Celis, M.~Jung, X.-Q.~Li, and A.~Pich, ``{Scalar contributions to $b\to c
  (u) \tau \nu$ transitions},''
  \href{https://dx.doi.org/10.1016/j.physletb.2017.05.037}{Phys.\  Lett.\  B
  {\bfseries 771} (2017) 168--179} {\ttfamily
  [\href{https://arxiv.org/abs/1612.07757}{arXiv:1612.07757}]}.

\bibitem{Akeroyd:2017mhr}
A.~G.~Akeroyd and C.-H.~Chen, ``{Constraint on the branching ratio of $B_c \to
  \tau \bar{\nu}$ from LEP1 and consequences for $R(D^{(*)})$ anomaly},''
  \href{https://dx.doi.org/10.1103/PhysRevD.96.075011}{Phys.\  Rev.\  D
  {\bfseries 96} (2017) 075011} {\ttfamily
  [\href{https://arxiv.org/abs/1708.04072}{arXiv:1708.04072}]}.

\bibitem{Blanke:2018yud}
M.~Blanke, { et al.}, ``{Impact of polarization observables and $ B_c\to \tau
  \nu$ on new physics explanations of the $b\to c \tau \nu$ anomaly},''
  \href{https://dx.doi.org/10.1103/PhysRevD.99.075006}{Phys.\  Rev.\  D
  {\bfseries 99} (2019) 075006} {\ttfamily
  [\href{https://arxiv.org/abs/1811.09603}{arXiv:1811.09603}]}.

\bibitem{Aebischer:2021ilm}
J.~Aebischer and B.~Grinstein, ``{Standard Model prediction of the $B_{c}$
  lifetime},'' \href{https://dx.doi.org/10.1007/JHEP07(2021)130}{JHEP
  {\bfseries 07} (2021) 130} {\ttfamily
  [\href{https://arxiv.org/abs/2105.02988}{arXiv:2105.02988}]}.

\bibitem{Tanaka:2012nw}
M.~Tanaka and R.~Watanabe, ``{New physics in the weak interaction of $\bar B\to
  D^{(*)}\tau\bar\nu$},''
  \href{https://dx.doi.org/10.1103/PhysRevD.87.034028}{Phys.\  Rev.\  D
  {\bfseries 87} (2013) 034028} {\ttfamily
  [\href{https://arxiv.org/abs/1212.1878}{arXiv:1212.1878}]}.

\bibitem{Belle:2019ewo}
{\bfseries Belle} Collaboration, ``{Measurement of the $D^{\ast-}$ polarization
  in the decay $B^0 \to D^{\ast -}\tau^+\nu_{\tau}$}.'' {\ttfamily
  \href{https://arxiv.org/abs/1903.03102}{arXiv:1903.03102}}.

\bibitem{Crivellin:2012ye}
A.~Crivellin, C.~Greub, and A.~Kokulu, ``{Explaining $B\to D\tau\nu$, $B\to
  D^*\tau\nu$ and $B\to \tau\nu$ in a 2HDM of type III},''
  \href{https://dx.doi.org/10.1103/PhysRevD.86.054014}{Phys.\  Rev.\  D
  {\bfseries 86} (2012) 054014} {\ttfamily
  [\href{https://arxiv.org/abs/1206.2634}{arXiv:1206.2634}]}.

\bibitem{Ko:2012sv}
P.~Ko, Y.~Omura, and C.~Yu, ``{$B \to D^(*) \tau \nu$ and $B \to \tau \nu$ in
  chiral $U(1)^\prime$ models with flavored multi Higgs doublets},''
  \href{https://dx.doi.org/10.1007/JHEP03(2013)151}{JHEP {\bfseries 03} (2013)
  151} {\ttfamily [\href{https://arxiv.org/abs/1212.4607}{arXiv:1212.4607}]}.

\bibitem{Crivellin:2013wna}
A.~Crivellin, A.~Kokulu, and C.~Greub, ``{Flavor-phenomenology of
  two-Higgs-doublet models with generic Yukawa structure},''
  \href{https://dx.doi.org/10.1103/PhysRevD.87.094031}{Phys.\  Rev.\  D
  {\bfseries 87} (2013) 094031} {\ttfamily
  [\href{https://arxiv.org/abs/1303.5877}{arXiv:1303.5877}]}.

\bibitem{Cline:2015lqp}
J.~M.~Cline, ``{Scalar doublet models confront \ensuremath{\tau} and b
  anomalies},'' \href{https://dx.doi.org/10.1103/PhysRevD.93.075017}{Phys.\
  Rev.\  D {\bfseries 93} (2016) 075017} {\ttfamily
  [\href{https://arxiv.org/abs/1512.02210}{arXiv:1512.02210}]}.

\bibitem{Crivellin:2015hha}
A.~Crivellin, J.~Heeck, and P.~Stoffer, ``{A perturbed lepton-specific
  two-Higgs-doublet model facing experimental hints for physics beyond the
  Standard Model},''
  \href{https://dx.doi.org/10.1103/PhysRevLett.116.081801}{Phys.\  Rev.\
  Lett.\  {\bfseries 116} (2016) 081801} {\ttfamily
  [\href{https://arxiv.org/abs/1507.07567}{arXiv:1507.07567}]}.

\bibitem{Lee:2017kbi}
J.-P.~Lee, ``{$B\to D^{(*)}\tau\nu_\tau$ in the 2HDM with an anomalous $\tau$
  coupling},'' \href{https://dx.doi.org/10.1103/PhysRevD.96.055005}{Phys.\
  Rev.\  D {\bfseries 96} (2017) 055005} {\ttfamily
  [\href{https://arxiv.org/abs/1705.02465}{arXiv:1705.02465}]}.

\bibitem{Iguro:2017ysu}
S.~Iguro and K.~Tobe, ``{$R(D^{(*)})$ in a general two Higgs doublet model},''
  \href{https://dx.doi.org/10.1016/j.nuclphysb.2017.10.014}{Nucl.\  Phys.\  B
  {\bfseries 925} (2017) 560--606} {\ttfamily
  [\href{https://arxiv.org/abs/1708.06176}{arXiv:1708.06176}]}.

\bibitem{Iguro:2018qzf}
S.~Iguro and Y.~Omura, ``{Status of the semileptonic $B$ decays and muon g-2 in
  general 2HDMs with right-handed neutrinos},''
  \href{https://dx.doi.org/10.1007/JHEP05(2018)173}{JHEP {\bfseries 05} (2018)
  173} {\ttfamily [\href{https://arxiv.org/abs/1802.01732}{arXiv:1802.01732}]}.

\bibitem{Martinez:2018ynq}
R.~Martinez, C.~F.~Sierra, and G.~Valencia, ``{Beyond $\mathcal{R}(D^{(*)})$
  with the general type-III 2HDM for $b\to c\tau\nu$},''
  \href{https://dx.doi.org/10.1103/PhysRevD.98.115012}{Phys.\  Rev.\  D
  {\bfseries 98} (2018) 115012} {\ttfamily
  [\href{https://arxiv.org/abs/1805.04098}{arXiv:1805.04098}]}.

\bibitem{Fraser:2018aqj}
S.~Fraser, C.~Marzo, L.~Marzola, M.~Raidal, and C.~Spethmann, ``{Towards a
  viable scalar interpretation of $R_{D^{(*)}}$},''
  \href{https://dx.doi.org/10.1103/PhysRevD.98.035016}{Phys.\  Rev.\  D
  {\bfseries 98} (2018) 035016} {\ttfamily
  [\href{https://arxiv.org/abs/1805.08189}{arXiv:1805.08189}]}.

\bibitem{Athron:2021auq}
P.~Athron, { et al.}, ``{Likelihood analysis of the flavour anomalies and g
  \textendash{} 2 in the general two Higgs doublet model},''
  \href{https://dx.doi.org/10.1007/JHEP01(2022)037}{JHEP {\bfseries 01} (2022)
  037} {\ttfamily [\href{https://arxiv.org/abs/2111.10464}{arXiv:2111.10464}]}.

\bibitem{Iguro:2022uzz}
S.~Iguro, ``{Revival of H- interpretation of RD(*) anomaly and closing low mass
  window},'' \href{https://dx.doi.org/10.1103/PhysRevD.105.095011}{Phys.\
  Rev.\  D {\bfseries 105} (2022) 095011} {\ttfamily
  [\href{https://arxiv.org/abs/2201.06565}{arXiv:2201.06565}]}.

\bibitem{Blanke:2022pjy}
M.~Blanke, S.~Iguro, and H.~Zhang, ``{Towards ruling out the charged Higgs
  interpretation of the $ {R}_{D^{\left(\ast \right)}} $ anomaly},''
  \href{https://dx.doi.org/10.1007/JHEP06(2022)043}{JHEP {\bfseries 06} (2022)
  043} {\ttfamily [\href{https://arxiv.org/abs/2202.10468}{arXiv:2202.10468}]}.

\bibitem{Kumar:2022rcf}
G.~Kumar, ``{Interplay of the charged Higgs boson effects in $R_{D^{(*)}}$, $b
  \rightarrow s \ell^{+}\ell^{-}$, and $W$ mass},''
  \href{https://dx.doi.org/10.1103/PhysRevD.107.075016}{Phys.\  Rev.\  D
  {\bfseries 107} (2023) 075016} {\ttfamily
  [\href{https://arxiv.org/abs/2212.07233}{arXiv:2212.07233}]}.

\bibitem{Iguro:2023jju}
S.~Iguro, ``{Conclusive probe of the charged Higgs solution of $P_5^\prime$ and
  $R_{D^{(*)}}$ discrepancies},''
  \href{https://dx.doi.org/10.1103/PhysRevD.107.095004}{Phys.\  Rev.\  D
  {\bfseries 107} (2023) 095004} {\ttfamily
  [\href{https://arxiv.org/abs/2302.08935}{arXiv:2302.08935}]}.

\bibitem{Das:2023gfz}
N.~Das, A.~Adhikary, and R.~Dutta, ``{Revisiting $b \to c\tau\nu$ anomalies
  with charged Higgs boson}.'' {\ttfamily
  \href{https://arxiv.org/abs/2305.17766}{arXiv:2305.17766}}.

\bibitem{Angelescu:2018tyl}
A.~Angelescu, D.~Be\v{c}irevi\'c, D.~A.~Faroughy, and O.~Sumensari, ``{Closing
  the window on single leptoquark solutions to the $B$-physics anomalies},''
  \href{https://dx.doi.org/10.1007/JHEP10(2018)183}{JHEP {\bfseries 10} (2018)
  183} {\ttfamily [\href{https://arxiv.org/abs/1808.08179}{arXiv:1808.08179}]}.

\bibitem{Kosnik:2012dj}
N.~Kosnik, ``{Model independent constraints on leptoquarks from $b \to s \ell^+
  \ell^-$ processes},''
  \href{https://dx.doi.org/10.1103/PhysRevD.86.055004}{Phys.\  Rev.\  D
  {\bfseries 86} (2012) 055004} {\ttfamily
  [\href{https://arxiv.org/abs/1206.2970}{arXiv:1206.2970}]}.

\bibitem{Shaw:2018sbe}
A.~Shaw, A.~Biswas, and A.~K.~Swain, ``{Collider signature of \(V_2\)
  Leptoquark with \( b\to s\) flavour observables},''
  \href{https://dx.doi.org/10.31526/lhep.2.2019.126}{LHEP {\bfseries 2} (2019)
  126} {\ttfamily [\href{https://arxiv.org/abs/1811.08887}{arXiv:1811.08887}]}.

\bibitem{Cheung:2022zsb}
K.~Cheung, W.-Y.~Keung, and P.-Y.~Tseng, ``{Isodoublet vector leptoquark
  solution to the muon g-2, RK,K*, RD,D*, and W-mass anomalies},''
  \href{https://dx.doi.org/10.1103/PhysRevD.106.015029}{Phys.\  Rev.\  D
  {\bfseries 106} (2022) 015029} {\ttfamily
  [\href{https://arxiv.org/abs/2204.05942}{arXiv:2204.05942}]}.

\bibitem{Sakaki:2013bfa}
Y.~Sakaki, M.~Tanaka, A.~Tayduganov, and R.~Watanabe, ``{Testing leptoquark
  models in $\bar B \to D^{(*)} \tau \bar\nu$},''
  \href{https://dx.doi.org/10.1103/PhysRevD.88.094012}{Phys.\  Rev.\  D
  {\bfseries 88} (2013) 094012} {\ttfamily
  [\href{https://arxiv.org/abs/1309.0301}{arXiv:1309.0301}]}.

\bibitem{Barbieri:1995uv}
R.~Barbieri, G.~R.~Dvali, and L.~J.~Hall, ``{Predictions from a U(2) flavor
  symmetry in supersymmetric theories},''
  \href{https://dx.doi.org/10.1016/0370-2693(96)00318-8}{Phys.\  Lett.\  B
  {\bfseries 377} (1996) 76--82} {\ttfamily
  [\href{https://arxiv.org/abs/hep-ph/9512388}{hep-ph/9512388}]}.

\bibitem{Barbieri:1997tu}
R.~Barbieri, L.~J.~Hall, and A.~Romanino, ``{Consequences of a U(2) flavor
  symmetry},'' \href{https://dx.doi.org/10.1016/S0370-2693(97)00372-9}{Phys.\
  Lett.\  B {\bfseries 401} (1997) 47--53} {\ttfamily
  [\href{https://arxiv.org/abs/hep-ph/9702315}{hep-ph/9702315}]}.

\bibitem{Barbieri:2011ci}
R.~Barbieri, G.~Isidori, J.~Jones-Perez, P.~Lodone, and D.~M.~Straub, ``{$U(2)$
  and Minimal Flavour Violation in Supersymmetry},''
  \href{https://dx.doi.org/10.1140/epjc/s10052-011-1725-z}{Eur.\  Phys.\  J.\
  C {\bfseries 71} (2011) 1725} {\ttfamily
  [\href{https://arxiv.org/abs/1105.2296}{arXiv:1105.2296}]}.

\bibitem{Barbieri:2011fc}
R.~Barbieri, P.~Campli, G.~Isidori, F.~Sala, and D.~M.~Straub, ``{$B$-decay
  CP-asymmetries in SUSY with a $U(2)^3$ flavour symmetry},''
  \href{https://dx.doi.org/10.1140/epjc/s10052-011-1812-1}{Eur.\  Phys.\  J.\
  C {\bfseries 71} (2011) 1812} {\ttfamily
  [\href{https://arxiv.org/abs/1108.5125}{arXiv:1108.5125}]}.

\bibitem{Barbieri:2012uh}
R.~Barbieri, D.~Buttazzo, F.~Sala, and D.~M.~Straub, ``{Flavour physics from an
  approximate $U(2)^3$ symmetry},''
  \href{https://dx.doi.org/10.1007/JHEP07(2012)181}{JHEP {\bfseries 07} (2012)
  181} {\ttfamily [\href{https://arxiv.org/abs/1203.4218}{arXiv:1203.4218}]}.

\bibitem{Blankenburg:2012nx}
G.~Blankenburg, G.~Isidori, and J.~Jones-Perez, ``{Neutrino Masses and LFV from
  Minimal Breaking of $U(3)^5$ and $U(2)^5$ flavor Symmetries},''
  \href{https://dx.doi.org/10.1140/epjc/s10052-012-2126-7}{Eur.\  Phys.\  J.\
  C {\bfseries 72} (2012) 2126} {\ttfamily
  [\href{https://arxiv.org/abs/1204.0688}{arXiv:1204.0688}]}.

\bibitem{Barbieri:2015yvd}
R.~Barbieri, G.~Isidori, A.~Pattori, and F.~Senia, ``{Anomalies in $B$-decays
  and $U(2)$ flavour symmetry},''
  \href{https://dx.doi.org/10.1140/epjc/s10052-016-3905-3}{Eur.\  Phys.\  J.\
  C {\bfseries 76} (2016) 67} {\ttfamily
  [\href{https://arxiv.org/abs/1512.01560}{arXiv:1512.01560}]}.

\bibitem{Bordone:2018nbg}
M.~Bordone, C.~Cornella, J.~Fuentes-Mart\'\i{}n, and G.~Isidori, ``{Low-energy
  signatures of the $\mathrm{PS}^3$ model: from $B$-physics anomalies to
  LFV},'' \href{https://dx.doi.org/10.1007/JHEP10(2018)148}{JHEP {\bfseries 10}
  (2018) 148} {\ttfamily
  [\href{https://arxiv.org/abs/1805.09328}{arXiv:1805.09328}]}.

\bibitem{Cornella:2019hct}
C.~Cornella, J.~Fuentes-Martin, and G.~Isidori, ``{Revisiting the vector
  leptoquark explanation of the B-physics anomalies},''
  \href{https://dx.doi.org/10.1007/JHEP07(2019)168}{JHEP {\bfseries 07} (2019)
  168} {\ttfamily [\href{https://arxiv.org/abs/1903.11517}{arXiv:1903.11517}]}.

\bibitem{Fuentes-Martin:2019mun}
J.~Fuentes-Mart\'\i{}n, G.~Isidori, J.~Pag\`es, and K.~Yamamoto, ``{With or
  without U(2)? Probing non-standard flavor and helicity structures in
  semileptonic B decays},''
  \href{https://dx.doi.org/10.1016/j.physletb.2019.135080}{Phys.\  Lett.\  B
  {\bfseries 800} (2020) 135080} {\ttfamily
  [\href{https://arxiv.org/abs/1909.02519}{arXiv:1909.02519}]}.

\bibitem{Cornella:2021sby}
C.~Cornella, D.~A.~Faroughy, J.~Fuentes-Martin, G.~Isidori, and M.~Neubert,
  ``{Reading the footprints of the B-meson flavor anomalies},''
  \href{https://dx.doi.org/10.1007/JHEP08(2021)050}{JHEP {\bfseries 08} (2021)
  050} {\ttfamily [\href{https://arxiv.org/abs/2103.16558}{arXiv:2103.16558}]}.

\bibitem{FernandezNavarro:2022gst}
M.~Fern\'andez~Navarro and S.~F.~King, ``{$B$-anomalies in a twin Pati-Salam
  theory of flavour}.'' {\ttfamily
  \href{https://arxiv.org/abs/2209.00276}{arXiv:2209.00276}}.

\bibitem{Iguro:2018vqb}
S.~Iguro, T.~Kitahara, Y.~Omura, R.~Watanabe, and K.~Yamamoto, ``{$D^{*}$
  polarization vs. $ {R}_{D^{\left(\ast \right)}} $ anomalies in the leptoquark
  models},'' \href{https://dx.doi.org/10.1007/JHEP02(2019)194}{JHEP {\bfseries
  02} (2019) 194} {\ttfamily
  [\href{https://arxiv.org/abs/1811.08899}{arXiv:1811.08899}]}.

\bibitem{Capdevila:2017iqn}
B.~Capdevila, A.~Crivellin, S.~Descotes-Genon, L.~Hofer, and J.~Matias,
  ``{Searching for New Physics with $b\to s\tau^+\tau^-$ processes},''
  \href{https://dx.doi.org/10.1103/PhysRevLett.120.181802}{Phys.\  Rev.\
  Lett.\  {\bfseries 120} (2018) 181802} {\ttfamily
  [\href{https://arxiv.org/abs/1712.01919}{arXiv:1712.01919}]}.

\bibitem{Allwicher:2022mcg}
L.~Allwicher, D.~A.~Faroughy, F.~Jaffredo, O.~Sumensari, and F.~Wilsch,
  ``{HighPT: A tool for high-pT Drell-Yan tails beyond the standard model},''
  \href{https://dx.doi.org/10.1016/j.cpc.2023.108749}{Comput.\  Phys.\
  Commun.\  {\bfseries 289} (2023) 108749} {\ttfamily
  [\href{https://arxiv.org/abs/2207.10756}{arXiv:2207.10756}]}.

\bibitem{Frampton:1989fu}
P.~H.~Frampton and B.-H.~Lee, ``{SU(15) GRAND UNIFICATION},''
  \href{https://dx.doi.org/10.1103/PhysRevLett.64.619}{Phys.\  Rev.\  Lett.\
  {\bfseries 64} (1990) 619}.

\bibitem{Frampton:1990hz}
P.~H.~Frampton and T.~W.~Kephart, ``{Higgs sector and proton decay in SU(15)
  grand unification},''
  \href{https://dx.doi.org/10.1103/PhysRevD.42.3892}{Phys.\  Rev.\  D
  {\bfseries 42} (1990) 3892--3894}.

\bibitem{Frampton:1991ay}
P.~H.~Frampton, ``{Light leptoquarks as possible signature of strong
  electroweak unification},''
  \href{https://dx.doi.org/10.1142/S0217732392000525}{Mod.\  Phys.\  Lett.\  A
  {\bfseries 7} (1992) 559--562}.

\bibitem{Bernigaud:2021fwn}
J.~Bernigaud, M.~Blanke, I.~de~Medeiros~Varzielas, J.~Talbert, and J.~Zurita,
  ``{LHC signatures of \ensuremath{\tau}-flavoured vector leptoquarks},''
  \href{https://dx.doi.org/10.1007/JHEP08(2022)127}{JHEP {\bfseries 08} (2022)
  127} {\ttfamily [\href{https://arxiv.org/abs/2112.12129}{arXiv:2112.12129}]}.

\bibitem{Cabibbo:1963yz}
N.~Cabibbo, ``{Unitary Symmetry and Leptonic Decays},''
  \href{https://dx.doi.org/10.1103/PhysRevLett.10.531}{Phys.\  Rev.\  Lett.\
  {\bfseries 10} (1963) 531--533}.

\bibitem{Kobayashi:1973fv}
M.~Kobayashi and T.~Maskawa, ``{CP Violation in the Renormalizable Theory of
  Weak Interaction},'' \href{https://dx.doi.org/10.1143/PTP.49.652}{Prog.\
  Theor.\  Phys.\  {\bfseries 49} (1973) 652--657}.

\bibitem{Kim:2018oih}
T.~J.~Kim, P.~Ko, J.~Li, J.~Park, and P.~Wu, ``{Correlation between $
  {R}_{D^{\left(\ast \right)}} $ and top quark FCNC decays in leptoquark
  models},'' \href{https://dx.doi.org/10.1007/JHEP07(2019)025}{JHEP {\bfseries
  07} (2019) 025} {\ttfamily
  [\href{https://arxiv.org/abs/1812.08484}{arXiv:1812.08484}]}.

\bibitem{Wolfenstein:1983yz}
L.~Wolfenstein, ``{Parametrization of the Kobayashi-Maskawa Matrix},''
  \href{https://dx.doi.org/10.1103/PhysRevLett.51.1945}{Phys.\  Rev.\  Lett.\
  {\bfseries 51} (1983) 1945}.

\bibitem{Jenkins:2013wua}
E.~E.~Jenkins, A.~V.~Manohar, and M.~Trott, ``{Renormalization Group Evolution
  of the Standard Model Dimension Six Operators II: Yukawa Dependence},''
  \href{https://dx.doi.org/10.1007/JHEP01(2014)035}{JHEP {\bfseries 01} (2014)
  035} {\ttfamily [\href{https://arxiv.org/abs/1310.4838}{arXiv:1310.4838}]}.

\bibitem{Alonso:2013hga}
R.~Alonso, E.~E.~Jenkins, A.~V.~Manohar, and M.~Trott, ``{Renormalization Group
  Evolution of the Standard Model Dimension Six Operators III: Gauge Coupling
  Dependence and Phenomenology},''
  \href{https://dx.doi.org/10.1007/JHEP04(2014)159}{JHEP {\bfseries 04} (2014)
  159} {\ttfamily [\href{https://arxiv.org/abs/1312.2014}{arXiv:1312.2014}]}.

\bibitem{Gonzalez-Alonso:2017iyc}
M.~Gonz\'alez-Alonso, J.~Martin~Camalich, and K.~Mimouni,
  ``{Renormalization-group evolution of new physics contributions to
  (semi)leptonic meson decays},''
  \href{https://dx.doi.org/10.1016/j.physletb.2017.07.003}{Phys.\  Lett.\  B
  {\bfseries 772} (2017) 777--785} {\ttfamily
  [\href{https://arxiv.org/abs/1706.00410}{arXiv:1706.00410}]}.

\bibitem{Aebischer:2017gaw}
J.~Aebischer, M.~Fael, C.~Greub, and J.~Virto, ``{B physics Beyond the Standard
  Model at One Loop: Complete Renormalization Group Evolution below the
  Electroweak Scale},'' \href{https://dx.doi.org/10.1007/JHEP09(2017)158}{JHEP
  {\bfseries 09} (2017) 158} {\ttfamily
  [\href{https://arxiv.org/abs/1704.06639}{arXiv:1704.06639}]}.

\bibitem{Aebischer:2018acj}
J.~Aebischer, A.~Crivellin, and C.~Greub, ``{QCD improved matching for
  semileptonic B decays with leptoquarks},''
  \href{https://dx.doi.org/10.1103/PhysRevD.99.055002}{Phys.\  Rev.\  D
  {\bfseries 99} (2019) 055002} {\ttfamily
  [\href{https://arxiv.org/abs/1811.08907}{arXiv:1811.08907}]}.

\bibitem{Bobeth:1999mk}
C.~Bobeth, M.~Misiak, and J.~Urban, ``{Photonic penguins at two loops and $m_t$
  dependence of $BR[B \to X_s l^+ l^-]$},''
  \href{https://dx.doi.org/10.1016/S0550-3213(00)00007-9}{Nucl.\  Phys.\  B
  {\bfseries 574} (2000) 291--330} {\ttfamily
  [\href{https://arxiv.org/abs/hep-ph/9910220}{hep-ph/9910220}]}.

\bibitem{Huber:2005ig}
T.~Huber, E.~Lunghi, M.~Misiak, and D.~Wyler, ``{Electromagnetic logarithms in
  $\bar B \to X_s l^+ l^-$},''
  \href{https://dx.doi.org/10.1016/j.nuclphysb.2006.01.037}{Nucl.\  Phys.\  B
  {\bfseries 740} (2006) 105--137} {\ttfamily
  [\href{https://arxiv.org/abs/hep-ph/0512066}{hep-ph/0512066}]}.

\bibitem{LHCb:2017myy}
{\bfseries LHCb} Collaboration, ``{Search for the decays $B_s^0\to\tau^+\tau^-$
  and $B^0\to\tau^+\tau^-$},''
  \href{https://dx.doi.org/10.1103/PhysRevLett.118.251802}{Phys.\  Rev.\
  Lett.\  {\bfseries 118} (2017) 251802} {\ttfamily
  [\href{https://arxiv.org/abs/1703.02508}{arXiv:1703.02508}]}.

\bibitem{LHCb:2018roe}
{\bfseries LHCb} Collaboration, ``{Physics case for an LHCb Upgrade II -
  Opportunities in flavour physics, and beyond, in the HL-LHC era}.''
  {\ttfamily \href{https://arxiv.org/abs/1808.08865}{arXiv:1808.08865}}.

\bibitem{BaBar:2016wgb}
{\bfseries BaBar} Collaboration, ``{Search for $B^{+}\rightarrow K^{+}
  \tau^{+}\tau^{-}$ at the BaBar experiment},''
  \href{https://dx.doi.org/10.1103/PhysRevLett.118.031802}{Phys.\  Rev.\
  Lett.\  {\bfseries 118} (2017) 031802} {\ttfamily
  [\href{https://arxiv.org/abs/1605.09637}{arXiv:1605.09637}]}.

\bibitem{Kou:2018nap}
E.~Kou and P.~Urquijo, eds., ``{The Belle II Physics Book},''
  \href{https://dx.doi.org/10.1093/ptep/ptz106}{PTEP {\bfseries 2019} (2019)
  123C01} {\ttfamily
  [\href{https://arxiv.org/abs/1808.10567}{arXiv:1808.10567}]}. [Erratum: PTEP
  2020, 029201 (2020)].

\bibitem{MILC:2015uhg}
{\bfseries MILC} Collaboration,
  ``{B\textrightarrow{}D\ensuremath{\ell}\ensuremath{\nu} form factors at
  nonzero recoil and |V$_{cb}$| from 2+1-flavor lattice QCD},''
  \href{https://dx.doi.org/10.1103/PhysRevD.92.034506}{Phys.\  Rev.\  D
  {\bfseries 92} (2015) 034506} {\ttfamily
  [\href{https://arxiv.org/abs/1503.07237}{arXiv:1503.07237}]}.

\bibitem{Belle:2017rcc}
{\bfseries Belle} Collaboration, ``{Precise determination of the CKM matrix
  element $\left| V_{cb}\right|$ with $\bar B^0 \to D^{*\,+} \, \ell^- \, \bar
  \nu_\ell$ decays with hadronic tagging at Belle}.'' {\ttfamily
  \href{https://arxiv.org/abs/1702.01521}{arXiv:1702.01521}}.

\bibitem{Belle:2018ezy}
{\bfseries Belle} Collaboration, ``{Measurement of the CKM matrix element
  $|V_{cb}|$ from $B^0\to D^{*-}\ell^ {+} \nu_\ell$ at Belle},''
  \href{https://dx.doi.org/10.1103/PhysRevD.100.052007}{Phys.\  Rev.\  D
  {\bfseries 100} (2019) 052007} {\ttfamily
  [\href{https://arxiv.org/abs/1809.03290}{arXiv:1809.03290}]}. [Erratum:
  Phys.Rev.D 103, 079901 (2021)].

\bibitem{Zheng:2020ult}
T.~Zheng, { et al.}, ``{Analysis of $B_c \to \tau\nu_\tau$ at CEPC},''
  \href{https://dx.doi.org/10.1088/1674-1137/abcf1f}{Chin.\  Phys.\  C
  {\bfseries 45} (2021) 023001} {\ttfamily
  [\href{https://arxiv.org/abs/2007.08234}{arXiv:2007.08234}]}.

\bibitem{Amhis:2021cfy}
Y.~Amhis, M.~Hartmann, C.~Helsens, D.~Hill, and O.~Sumensari, ``{Prospects for
  $ {B}_c^{+} $\textrightarrow{}
  \ensuremath{\tau}$^{+}$\ensuremath{\nu}$_{\tau}$ at FCC-ee},''
  \href{https://dx.doi.org/10.1007/JHEP12(2021)133}{JHEP {\bfseries 12} (2021)
  133} {\ttfamily [\href{https://arxiv.org/abs/2105.13330}{arXiv:2105.13330}]}.

\bibitem{Fedele:2023gyi}
M.~Fedele, { et al.}, ``{Prospects for $B_c^+$ and $B^+\to \tau^+ \nu_\tau$ at
  FCC-ee}.'' {\ttfamily
  \href{https://arxiv.org/abs/2305.02998}{arXiv:2305.02998}}.

\bibitem{Sirunyan:2018vhk}
{\bfseries CMS} Collaboration, ``{Search for heavy neutrinos and
  third-generation leptoquarks in hadronic states of two $\tau$ leptons and two
  jets in proton-proton collisions at $\sqrt{s} =$ 13 TeV},''
  \href{https://dx.doi.org/10.1007/JHEP03(2019)170}{JHEP {\bfseries 03} (2019)
  170} {\ttfamily [\href{https://arxiv.org/abs/1811.00806}{arXiv:1811.00806}]}.

\bibitem{Aaboud:2019bye}
{\bfseries ATLAS} Collaboration, ``{Searches for third-generation scalar
  leptoquarks in $\sqrt{s}$ = 13 TeV pp collisions with the ATLAS detector},''
  \href{https://dx.doi.org/10.1007/JHEP06(2019)144}{JHEP {\bfseries 06} (2019)
  144} {\ttfamily [\href{https://arxiv.org/abs/1902.08103}{arXiv:1902.08103}]}.

\bibitem{Aad:2021rrh}
{\bfseries ATLAS} Collaboration, ``{Search for pair production of
  third-generation scalar leptoquarks decaying into a top quark and a
  $\tau$-lepton in $pp$ collisions at $ \sqrt{s} $ = 13 TeV with the ATLAS
  detector},'' \href{https://dx.doi.org/10.1007/JHEP06(2021)179}{JHEP
  {\bfseries 06} (2021) 179} {\ttfamily
  [\href{https://arxiv.org/abs/2101.11582}{arXiv:2101.11582}]}.

\bibitem{ATLAS:2021jyv}
{\bfseries ATLAS} Collaboration, ``{Search for new phenomena in $pp$ collisions
  in final states with tau leptons, b-jets, and missing transverse momentum
  with the ATLAS detector},''
  \href{https://dx.doi.org/10.1103/PhysRevD.104.112005}{Phys.\  Rev.\  D
  {\bfseries 104} (2021) 112005} {\ttfamily
  [\href{https://arxiv.org/abs/2108.07665}{arXiv:2108.07665}]}.

\bibitem{Faroughy:2016osc}
D.~A.~Faroughy, A.~Greljo, and J.~F.~Kamenik, ``{Confronting lepton flavor
  universality violation in B decays with high-$p_T$ tau lepton searches at
  LHC},'' \href{https://dx.doi.org/10.1016/j.physletb.2016.11.011}{Phys.\
  Lett.\  B {\bfseries 764} (2017) 126--134} {\ttfamily
  [\href{https://arxiv.org/abs/1609.07138}{arXiv:1609.07138}]}.

\bibitem{Iguro:2018fni}
S.~Iguro, Y.~Omura, and M.~Takeuchi, ``{Test of the $R(D^{(*)})$ anomaly at the
  LHC},'' \href{https://dx.doi.org/10.1103/PhysRevD.99.075013}{Phys.\  Rev.\  D
  {\bfseries 99} (2019) 075013} {\ttfamily
  [\href{https://arxiv.org/abs/1810.05843}{arXiv:1810.05843}]}.

\bibitem{Mandal:2018kau}
T.~Mandal, S.~Mitra, and S.~Raz, ``{$R_{D^{(*)}}$ motivated $\mathcal{S}_1$
  leptoquark scenarios: Impact of interference on the exclusion limits from LHC
  data},'' \href{https://dx.doi.org/10.1103/PhysRevD.99.055028}{Phys.\  Rev.\
  D {\bfseries 99} (2019) 055028} {\ttfamily
  [\href{https://arxiv.org/abs/1811.03561}{arXiv:1811.03561}]}.

\bibitem{Greljo:2018tzh}
A.~Greljo, J.~Martin~Camalich, and J.~D.~Ruiz-\'Alvarez, ``{Mono-$\tau$
  Signatures at the LHC Constrain Explanations of $B$-decay Anomalies},''
  \href{https://dx.doi.org/10.1103/PhysRevLett.122.131803}{Phys.\  Rev.\
  Lett.\  {\bfseries 122} (2019) 131803} {\ttfamily
  [\href{https://arxiv.org/abs/1811.07920}{arXiv:1811.07920}]}.

\bibitem{Altmannshofer:2017poe}
W.~Altmannshofer, P.~S.~Bhupal~Dev, and A.~Soni, ``{$R_{D^{(*)}}$ anomaly: A
  possible hint for natural supersymmetry with $R$-parity violation},''
  \href{https://dx.doi.org/10.1103/PhysRevD.96.095010}{Phys.\  Rev.\  D
  {\bfseries 96} (2017) 095010} {\ttfamily
  [\href{https://arxiv.org/abs/1704.06659}{arXiv:1704.06659}]}.

\bibitem{Abdullah:2018ets}
M.~Abdullah, J.~Calle, B.~Dutta, A.~Fl\'orez, and D.~Restrepo, ``{Probing a
  simplified, $W^{\prime}$ model of $R(D^{(\ast)})$ anomalies using $b$-tags,
  $\tau$ leptons and missing energy},''
  \href{https://dx.doi.org/10.1103/PhysRevD.98.055016}{Phys.\  Rev.\  D
  {\bfseries 98} (2018) 055016} {\ttfamily
  [\href{https://arxiv.org/abs/1805.01869}{arXiv:1805.01869}]}.

\bibitem{Marzocca:2020ueu}
D.~Marzocca, U.~Min, and M.~Son, ``{Bottom-Flavored Mono-Tau Tails at the
  LHC},'' \href{https://dx.doi.org/10.1007/JHEP12(2020)035}{JHEP {\bfseries 12}
  (2020) 035} {\ttfamily
  [\href{https://arxiv.org/abs/2008.07541}{arXiv:2008.07541}]}.

\bibitem{Bhaskar:2021pml}
A.~Bhaskar, D.~Das, T.~Mandal, S.~Mitra, and C.~Neeraj, ``{Precise limits on
  the charge-2/3 U1 vector leptoquark},''
  \href{https://dx.doi.org/10.1103/PhysRevD.104.035016}{Phys.\  Rev.\  D
  {\bfseries 104} (2021) 035016} {\ttfamily
  [\href{https://arxiv.org/abs/2101.12069}{arXiv:2101.12069}]}.

\bibitem{Iguro:2020keo}
S.~Iguro, M.~Takeuchi, and R.~Watanabe, ``{Testing Leptoquark/EFT in $\bar B
  \to D^{(*)} l\bar\nu$ at the LHC},''
  \href{https://dx.doi.org/10.1140/epjc/s10052-021-09125-5}{Eur.\  Phys.\  J.\
  C {\bfseries 81} (2021) 406} {\ttfamily
  [\href{https://arxiv.org/abs/2011.02486}{arXiv:2011.02486}]}.

\bibitem{Endo:2021lhi}
M.~Endo, S.~Iguro, T.~Kitahara, M.~Takeuchi, and R.~Watanabe, ``{Non-resonant
  new physics search at the LHC for the $b \to
  c\ensuremath{\tau}\ensuremath{\nu}$ anomalies},''
  \href{https://dx.doi.org/10.1007/JHEP02(2022)106}{JHEP {\bfseries 02} (2022)
  106} {\ttfamily [\href{https://arxiv.org/abs/2111.04748}{arXiv:2111.04748}]}.

\bibitem{CMS:2022goy}
{\bfseries CMS} Collaboration, ``{Searches for additional Higgs bosons and for
  vector leptoquarks in $\tau\tau$ final states in proton-proton collisions at
  $\sqrt{s}$ = 13 TeV}.'' {\ttfamily
  \href{https://arxiv.org/abs/2208.02717}{arXiv:2208.02717}}.

\bibitem{ATLAS:2020zms}
{\bfseries ATLAS} Collaboration, ``{Search for heavy Higgs bosons decaying into
  two tau leptons with the ATLAS detector using $pp$ collisions at
  $\sqrt{s}=13$ TeV},''
  \href{https://dx.doi.org/10.1103/PhysRevLett.125.051801}{Phys.\  Rev.\
  Lett.\  {\bfseries 125} (2020) 051801} {\ttfamily
  [\href{https://arxiv.org/abs/2002.12223}{arXiv:2002.12223}]}.

\bibitem{CMS:2015hmx}
{\bfseries CMS} Collaboration, ``{Search for W' decaying to tau lepton and
  neutrino in proton-proton collisions at $\sqrt{s} =$ 8 TeV},''
  \href{https://dx.doi.org/10.1016/j.physletb.2016.02.002}{Phys.\  Lett.\  B
  {\bfseries 755} (2016) 196--216} {\ttfamily
  [\href{https://arxiv.org/abs/1508.04308}{arXiv:1508.04308}]}.

\bibitem{Aaboud:2018vgh}
{\bfseries ATLAS} Collaboration, ``{Search for High-Mass Resonances Decaying to
  $\tau\nu$ in pp Collisions at $\sqrt{s}$=13 TeV with the ATLAS Detector},''
  \href{https://dx.doi.org/10.1103/PhysRevLett.120.161802}{Phys.\  Rev.\
  Lett.\  {\bfseries 120} (2018) 161802} {\ttfamily
  [\href{https://arxiv.org/abs/1801.06992}{arXiv:1801.06992}]}.

\bibitem{Sirunyan:2018lbg}
{\bfseries CMS} Collaboration, ``{Search for a W' boson decaying to a $\tau$
  lepton and a neutrino in proton-proton collisions at $\sqrt{s} =$ 13 TeV},''
  \href{https://dx.doi.org/10.1016/j.physletb.2019.01.069}{Phys.\  Lett.\  B
  {\bfseries 792} (2019) 107--131} {\ttfamily
  [\href{https://arxiv.org/abs/1807.11421}{arXiv:1807.11421}]}.

\bibitem{ATLAS:2021bjk}
{\bfseries ATLAS} Collaboration, ``{Search for high-mass resonances in final
  states with a tau lepton and missing transverse momentum with the ATLAS
  detector, ATLAS-CONF-2021-025}.''.

\bibitem{PDG2022}
{\bfseries Particle Data Group} Collaboration, ``{Review of Particle
  Physics},'' \href{https://dx.doi.org/10.1093/ptep/ptac097}{PTEP {\bfseries
  2022} (2022) 083C01}.

\bibitem{Tanaka:2016ijq}
M.~Tanaka and R.~Watanabe, ``{New physics contributions in $B\to\pi\tau\bar\nu$
  and $B\to\tau\bar\nu$},'' \href{https://dx.doi.org/10.1093/ptep/ptw175}{PTEP
  {\bfseries 2017} (2017) 013B05} {\ttfamily
  [\href{https://arxiv.org/abs/1608.05207}{arXiv:1608.05207}]}.

\bibitem{BESIII:2019vhn}
{\bfseries BESIII} Collaboration, ``{Observation of the leptonic decay $D^+ \to
  \tau^+ \nu_\tau$},''
  \href{https://dx.doi.org/10.1103/PhysRevLett.123.211802}{Phys.\  Rev.\
  Lett.\  {\bfseries 123} (2019) 211802} {\ttfamily
  [\href{https://arxiv.org/abs/1908.08877}{arXiv:1908.08877}]}.

\bibitem{Belle:2017oht}
{\bfseries Belle} Collaboration, ``{Search for $\boldsymbol{B\to
  h\nu\bar{\nu}}$ decays with semileptonic tagging at Belle},''
  \href{https://dx.doi.org/10.1103/PhysRevD.96.091101}{Phys.\  Rev.\  D
  {\bfseries 96} (2017) 091101} {\ttfamily
  [\href{https://arxiv.org/abs/1702.03224}{arXiv:1702.03224}]}. [Addendum:
  Phys.Rev.D 97, 099902 (2018)].

\bibitem{Buras:2014fpa}
A.~J.~Buras, J.~Girrbach-Noe, C.~Niehoff, and D.~M.~Straub, ``{$ B\to
  {K}^{\left(\ast \right)}\nu \overline{\nu} $ decays in the Standard Model and
  beyond},'' \href{https://dx.doi.org/10.1007/JHEP02(2015)184}{JHEP {\bfseries
  02} (2015) 184} {\ttfamily
  [\href{https://arxiv.org/abs/1409.4557}{arXiv:1409.4557}]}.

\bibitem{Bause:2022rrs}
R.~Bause, H.~Gisbert, M.~Golz, and G.~Hiller, ``{Model-independent analysis of
  $b\to d$ processes}.'' {\ttfamily
  \href{https://arxiv.org/abs/2209.04457}{arXiv:2209.04457}}.

\bibitem{Iguro:2022ozl}
S.~Iguro, J.~Kawamura, S.~Okawa, and Y.~Omura, ``{Importance of vector
  leptoquark-scalar box diagrams in Pati-Salam unification with vector-like
  families},'' \href{https://dx.doi.org/10.1007/JHEP07(2022)022}{JHEP
  {\bfseries 07} (2022) 022} {\ttfamily
  [\href{https://arxiv.org/abs/2201.04638}{arXiv:2201.04638}]}.

\bibitem{Iguro:2021kdw}
S.~Iguro, J.~Kawamura, S.~Okawa, and Y.~Omura, ``{TeV-scale vector leptoquark
  from Pati-Salam unification with vectorlike families},''
  \href{https://dx.doi.org/10.1103/PhysRevD.104.075008}{Phys.\  Rev.\  D
  {\bfseries 104} (2021) 075008} {\ttfamily
  [\href{https://arxiv.org/abs/2103.11889}{arXiv:2103.11889}]}.

\bibitem{IOprojectNo17}
S.~Iguro and Y.~Omura, ``work in progress.''.

\end{thebibliography}\endgroup
}
\end{document}